\documentclass[onefignum,onetabnum]{siamonline171218}

\usepackage{amssymb,amsfonts,amscd} 
\usepackage{commath}
\usepackage{setspace,geometry}
\usepackage{subcaption}
\usepackage{fullpage}
\usepackage[toc,page]{appendix}
\usepackage{tikz}
\usetikzlibrary{shapes,arrows}
\graphicspath{{figs/}}
\usepackage{verbatim}      	
\usepackage{makeidx}       	
\usepackage{mathtools}
\usepackage{color, bm, bbm, url, upgreek, gensymb}
\usepackage[version=3]{mhchem}
\usepackage{multicol, multirow} 
\usepackage{pdfpages}
\usepackage{cleveref}
\usepackage{rotating}
\usepackage{siunitx}
\usepackage{todonotes}

\newcommand\be{\begin{equation}}
\newcommand\ee{\end{equation}}
\newcommand\bal{\begin{align}}
\newcommand\eal{\end{align}}

\newcommand\revise[1]{{\color{black}#1}}

\newcommand\bd{\bm{d}}

\newcommand\h{\text{H}}

\newcommand\hho{\text{H}_2\text{O}}

\newcommand\commentout[1]{}

\title{Representing model inadequacy: A stochastic operator approach}

\author{Rebecca E. Morrison, Todd A. Oliver, Robert D. Moser}

\ifpdf
\hypersetup{
  pdftitle={Representing model inadequacy: A stochastic operator approach},
  pdfauthor={R. E. Morrison, T. A, Oliver, R. D. Moser}
}
\fi

\date{\today}
\begin{document}
\maketitle

\begin{abstract}
Mathematical models of physical systems are subject to many
uncertainties such as measurement errors and uncertain initial and
boundary conditions.  After accounting for these uncertainties, it is
often revealed that discrepancies between the model output and the
observations remain; if so, the model is said to be inadequate. In
practice, the inadequate model may be the best that is available or
tractable, and so it may be necessary to use the model for prediction
despite its inadequacy.  In this case, a representation of the
inadequacy is necessary, so the impact of the observed discrepancy can
be determined. We investigate this problem in the context of chemical
kinetics and propose a new technique to account for model inadequacy
that is both probabilistic and physically meaningful. A stochastic
inadequacy operator $\mathcal{S}$ is introduced which is embedded in
the ODEs describing the evolution of chemical species concentrations
and which respects certain physical constraints such as conservation
laws. The parameters of $\mathcal{S}$ are governed by probability
distributions, which in turn are characterized by a set of
hyperparameters. The model parameters and hyperparameters are
calibrated using high-dimensional hierarchical Bayesian inference. We
apply the method to a typical problem in chemical kinetics---the
reaction mechanism of hydrogen combustion.
\end{abstract}



\section{Introduction}\label{sec:intro}
Model inadequacy is a complex and critical issue that affects nearly all realms
of computational science and engineering. In general, models of physical systems
are imperfect: they rely on abstractions and simplifications which do not
perfectly represent the modeled system. Sometimes the imperfections are small
enough that any discrepancy between the model and reality is dominated by
observation error, such that the discrepancy is essentially undetectable given
existing measurement technology.  In contrast, a model is demonstrably
inadequate when the imperfections lead to a detectable inconsistency between the
model and observations.  Such inadequacies are often detected during model
validation, which is the process of assessing whether a given mathematical
model, including representations of relevant uncertainties, is consistent with
knowledge regarding the modeled system \cite{babuska08,bayarri07,Oliver_2015}.
When an inadequacy is detected, one would generally prefer to improve the model
to remove the discrepancy,
but such improvement is often not feasible.

\revise{One of the most important uses of computational models is to
  make predictions, by which we mean to predict values of model
  outputs without corroborating observations of the predicted
  quantities. Such predictions are important to engineering design and
  decision making. If one is using an inadequate model, as is often
  the case because that is all that is tractable, it is important to
  characterize the uncertainty in the prediction due to model errors
  \cite{Oliver_2015}. A representation of model inadequacy is
  therefore needed, and it is critical that the model inadequacy be
  represented in a way that will allow the uncertainty to be
  propagated to the predictions, as described by \cite{Oliver_2015}.}

\revise{Treatment of model inadequacy has been a topic of research in the Bayesian statistics
literature for a number of years.  A common approach is to pose and calibrate a purely statistical
representation of the discrepancy between the model output and the true value of that output (often called a
bias function)~\cite{kennedy01,bayarri07,higdon08,higdon04,swiler06}.  However, this representation
is unable to characterize the impact of inadequacy on predictions of an unobserved quantity
\cite{Oliver_2015}.}
\revise{Further, we generally have both qualitative and quantitative knowledge
about the phenomena being modeled, the reasons for the model inadequacy
and the way errors are injected into the model. It is important to
formulate the inadequacy formulation to respect this knowledge if
it is to reliably represent the uncertainty in the predictions.
Oliver {\it et al} \cite{Oliver_2015} developed such an inadequacy
representation for a trivial example, but broadly applicable techniques
for formulating inadequacy representations for models of complex
physical systems are not available.
}

\revise{We propose a new broadly applicable formulation for model
  inadequacy in terms of stochastic operators, which can satisfy the
  requirements above. Here, this approach is applied to models of
  chemical kinetics as an example of a relatively complex system with
  inadequate models and rich knowledge regarding the phenomenon.
Specifically, the stochastic model inadequacy operator is introduced
as a source in the equations describing the evolution of
the chemical system, and is constructed to
respect certain non-negotiable constraints on the system.}  This approach results in a model
inadequacy representation that is both probabilistic and physically meaningful. The formulation is
demonstrated on a typical problem in chemical kinetics, namely the reaction mechanism model for
hydrogen combustion, where it is shown that the formulation is flexible enough to account for
significant inadequacies present in the original model. 

Chemical mechanisms and kinetics models describe the process and rates
of chemical reactions \cite{steinfeld98, williams85}. In general, a
reaction mechanism is extraordinarily complex, even when there are
only two or three initial reactants. An accurate description of the
chemical processes involved in the oxidation of hydrocarbons, for
example, may include hundreds or thousands of reactions and fifty or
more chemical species \cite{grimech,westmoreland93}.  At the same
time, there is significant uncertainty in the reaction rates for these
reactions; recent efforts to address this include
\cite{konnov08,miki12,Braman_2013}.  Furthermore, kinetics models of
these chemical mechanisms must be embedded within a larger fluids
calculation to model practical combustion systems. The chemical
dynamics must then be represented at every point in space and
time. Because the computational cost of such detailed mechanisms is so
high, drastically reduced mechanisms are often used instead.  Such
reduced models are commonly found in the combustion literature
\cite{burke12, westmoreland93, williams08}.  However, errors
introduced by the reduced models may render the model inadequate even
if the detailed model it is based on is not.  Alternatively, it may be
that a highly accurate detailed reaction mechanism is not known, in
which case any available model should be viewed as a reduced mechanism
relative to the unknown reality.  This work is concerned with
accounting for the inadequacy resulting from the use of a reduced
chemical mechanism, though the representation proposed is equally applicable to inadequate
mechanisms.

If a chemical kinetics model is inadequate, it would be best to
improve the kinetics model directly to eliminate the
inadequacy. Indeed, refinement of chemical mechanisms in combustion is
an active topic of research; for a small sample focused on H$_2$/O$_2$
reactions, see \cite{burke11,burke12,conaire04,mueller99}. However,
this type of refinement is not always an option, because of a lack of
physical insight required to develop a higher fidelity model, a lack
of detailed observational data to support such model development, or a
lack of time or other resources required for the model development
process.  Further, even when a higher fidelity model exists, it may be
impractical to use due to a lack of the necessary computational
resources.  Thus, there is a general need for methods
that account for the discrepancy between the inadequate model and the
observations that do not require traditional model improvement.  To develop such
a representation, one must adopt a mathematical framework for reasoning about
the model inadequacy.  Here we adopt a Bayesian point of view
\cite{Cox_1961, jaynes03}, and thus, any lack of knowledge---and model
inadequacy specifically---is modeled using probability.  Further, the Bayesian
approach offers a natural framework for representing all uncertainties that
arise in using reduced kinetics models to make predictions---including modeling
inadequacy of course but also uncertain kinetics parameters and measurement
uncertainties---as well as a natural method for updating these representations
to incorporate information from data \cite{Christian_2001, Kaipo_2005,
  calvetti07, sivia06}.

In this work, inadequacy representations are formulated not as
corrections of the model output, but as stochastic enrichments of the
model itself which are specifically constructed to respect known
physical constraints.  A set of chemical reactions is modeled by
describing the time derivative of the species concentrations and
temperature, so the model consists of a set of nonlinear ordinary
differential equations. The necessary constraints in this context are
conservation of atoms, conservation of energy, and non-negativity of
concentrations.  The model inadequacy representation developed here is
characterized by a stochastic operator $\mathcal{S}$, which is added
to the formulation of the time derivatives of the species
concentrations and temperature.  This operator is constructed such
that all realizations of the solution of the enriched reduced
model---i.e. the reduced model coupled with the inadequacy
representation---conserve atoms and energy and have non-negative
concentrations at all times.  The main component of the operator is
additive, linear, and probabilistic, encoded in a random matrix $S$.
The use of the term {\it random matrix} implies that each entry is
characterized by a probability distribution.  This usage is consistent
with the definition of random matrices from random matrix theory (see
\cite{edelman05,mehta04}), although in that field a random matrix is
usually much less constrained than in the present case, and its
properties (such as the distributions of the eigenvalues) are found in
the limit as the size of the matrix goes to infinity. A few
applications of random matrices to engineering problems are presented
by Soize \cite{soize05,soize10}. However, this work also differs from
our approach in that the probability of a given matrix is
characterized by properties of the entire matrix, such as the
determinant, whereas we characterize each non-zero entry.  Moreover,
the inadequacy operator $\mathcal{S}$ is more general than a random
matrix alone, including two nonlinear operators as well.

%

\revise{Other stochastic formulations for chemically reacting systems have appeared in the
literature.  Of particular interest here is the approach pioneered by Gillespie~\cite{gillespie07stochastic}, in
which a stochastic model is formulated to represent aleatory uncertainty introduced by, for example,
quantum indeterminacy.  These ideas have found important applications in systems where the
populations of molecules involved is small so that the effects of these aleatory uncertainties can
lead to substantial deviations from classical deterministic models, such as gene regulatory
networks~\cite{karlebach2008modelling}.  However, 
  the goals of this approach are fundamentally different from those
  pursued here.  In particular, the Gillespie approach seeks to
  enhance the physical description of chemical kinetics to account for
  non-deterministic physical processes. In contrast, it is assumed here that the
  evolution of the species concentrations is in fact
  deterministic---i.e., that the species populations are sufficiently
  large that the stochastic effects represented by Gillespie are
  negligible---but that an entirely satisfactory deterministic model
  is either unavailable due to lack of knowledge or computationally
  intractable.
  
  Although the goals are quite different, the approach is similar in
  that the classical deterministic description of chemical kinetics is
  replaced by a stochastic description in which the species
  concentrations are random variables.  However, the form of stochasticity in typical
  stochastic kinetics is overly restrictive for the current purposes.
  For example, in a common stochastic kinetics approach, the model
  takes the form of the chemical Langevin equation, in which the
  deterministic kinetics model is modified by an additive white noise
  term.  In this case, the noise only modifies the source terms for
  species appearing in the original deterministic model.  When applied
  to the inadequate deterministic models considered here, this
  formulation cannot account for atoms that belong to species that are
  not tracked as part of the inadequate model and thus cannot fully
  capture the inadequacy.  In general, a richer representation of the
  inadequacy---one that accounts for both missing species and reaction
  pathways---is required.  Such a form is proposed in this work.}

The proposed form of the inadequacy representation satisfies
  all the constraints implied by conservation of atoms, conservation
  of energy, and non-negativity of concentrations. However, satisfying
  these constraints does not fully determine the stochastic inadequacy
  operator. The stochastic elements of the operator are described by
  hyperparameters that determine their mean and variance, and these
  can only be determined by reference to the actual discrepancy
  between the model and the real system, which is accomplished through
  a process of calibration. Further, just like any other model, the
  model enriched with its inadequacy representation must be validated
  against observations of the system.

  The calibration and validation of models of physical systems has
  been a subject of great interest in computational science and
  engineering
  \cite{oberkampf2004verification,schwer2009guide,oden2006revolutionizing}.
  Generally, calibration is an inverse problem in which model
  parameter values are inferred given data on outputs from the
  model. In the presence of uncertainty, such an inverse problem is
  naturally posed in terms of Bayesian inference. With
  uncertainties, the validation question of whether the model is
  consistent with the data is recast as a question whether it is
  improbable that the data could arise from the model, and a variety
  of statistical tests are available to measure that. The approach
  pursued here is essentially posterior predictive assessment as
  described by \cite{gelman96,rubin98} and by \cite{Moser_2016} for
  models of physical systems.
  These and similar
  calibration and validation techniques have been applied in a wide
  variety of applications, including turbulence
  \cite{OliverMoser2012,CheungEtal2011}, kinetics
  \cite{sargsyan2015statistical}, atomistic systems
  \cite{farrell2015bayesian}, flow in porous media
  \cite{barth2001predictive}, fatigue crack growth
  \cite{sankararaman2011uncertainty}, and cardio-vascular flows
  \cite{schiavazzi2017patient}, to name just a few.

Returning to the inadequacy representation considered here, the
hyperparameters characterizing the stochastic inadequacy operator need
to be calibrated as discussed above. Because the inadequacy operator
is stochastic, the forward problem mapping specific values of the
hyperparameters to the model outputs is also stochastic. This
complicates the Bayesian inverse problem. A hierarchical Bayesian
formulation is used to address this \cite{berliner96}.

%

The rest of the paper is organized as follows. In \S\ref{sec:kin}, we give a
brief overview of kinetics modeling. In \S\ref{sec:op}, the general formulation
and properties of the stochastic operator are presented.  \S\ref{sec:cal}
describes the Bayesian framework for calibration and validation of the various
models, including hierarchical Bayesian modeling and validation under
uncertainty. In \S\ref{sec:ex}, the approach is applied to the specific case of
hydrogen combustion.  Concluding remarks are given in \S\ref{sec:con}.

\section{Chemical mechanism models} \label{sec:kin}
Chemical mechanisms and kinetics models describe the process and rates of
chemical reactions. In a typical chemical reaction, there is a set of reactant
species which, after a complex series of intermediate reactions, ultimately form
the chemical products. These intermediate steps, in which chemical species react
directly with each other, are called elementary reactions. The set of elementary
reactions is called the reaction mechanism, and a typical combustion problem may
include tens to thousands of elementary reactions.  This section provides a
minimal introduction to the essentials of chemical kinetics necessary to
understand the development of the model inadequacy representation in
\S\ref{sec:op}.  For more details on chemical kinetics, see
\cite{steinfeld98} for a general text and \cite{williams85} for a presentation
focused on combustion.

To introduce the main concepts, consider the following example reaction set with
four species and two reversible 
elementary reactions:
\begin{equation*}
\cee{A + B <=>[k_1^f][k_1^b] C},\quad \quad \cee{A + C <=>[k_2^f][k_2^b] D},
\end{equation*}
where A, B, C, and D denote the different chemical species, and $k_1^f$, $k_2^f$
and $k_1^b$, $k_2^b$ are the forward and backward rate coefficients,
respectively.  Let $\bm{x} = [x_1, x_2, x_3, x_4]^T$ be the vector of molar
concentrations (having dimensions moles per unit volume) corresponding to
species A, B, C, D.
The rate of each reaction is often modeled as linear in the concentration of the reactants, although
this power, or order, associated with a given species may be non-unity. With the assumption of
linearity in each species, the forward rate expressions of the two reactions are thus
\begin{equation*}
r_1^f = k_1^f x_1 x_2,\quad \quad r_2^f = k_2^f x_1 x_3.
\end{equation*}
Similarly, the backward rates are given by
\begin{equation*}
r_1^b = k_1^b x_3,\quad \quad r_2^b = k_2^b x_4.
\end{equation*}
Finally, the ODEs for the molar concentrations are
\begin{align*} 
\dot{x}_1 &= -r_1^f + r_1^b - r_2^f + r_2^b\\ 
\dot{x}_2 &= -r_1^f + r_1^b \\ 
\dot{x}_3 &= +r_1^f - r_1^b - r_2^f + r_2^b \\
\dot{x}_4 &= +r_2^f - r_2^b .
\end{align*} 

The rate coefficients $k$ are generally functions of temperature, and may follow
a given empirical form depending on the specific reaction. A common form is the
Arrhenius Law, $k(T) = A e^{-E/(R^{\circ}T)}$,
for some prefactor $A$, activation energy $E$, and universal gas constant
$R^{\circ}$. Another common form is the modified Arrhenius,
$k(T) = A T^{b} e^{-E/(R^{\circ}T)}$,
with the additional constant $b$. These 
expressions are often used in the literature to describe the forward rate coefficient, and this is true in
the example problem in \S\ref{sec:ex}.  However, the backwards rate coefficients are usually not
specified.  Instead, these are determined from the equilibrium constant which depends only on the
thermodynamics of the reaction.  See~\cite{steinfeld98} for details.

Given a reversible reaction and the forward rate coefficients, there are various
software libraries which will solve for the backwards rate coefficients. In this
work, a chemistry software library called Antioch (A New Templated
Implementation of Chemistry for Hydrodynamics) was used to set up the chemical
model, query thermodynamic information, and solve for the reverse reaction rates
\cite{antioch}. 

To complete the specification of the system, a governing equation for
temperature is required.  This equation is derived based on
conservation of energy.  In this work, we consider a reacting mixture
of ideal gases.  Further, the reactions are assumed to occur in a
constant volume that does not exchange heat or mass with its
surroundings.  In this case, changes in the system temperature are due
only to the difference in chemical energy between the reactants and
products. For an ideal gas, the internal energy depends only
on temperature (not $v$ or $p$) and the species concentrations:
\begin{equation*}
U(T, \bm{x}) = \sum_i u_i(T) x_i.
\end{equation*}
Thus,
\begin{equation*}
\frac{dU}{dt} = \sum_i{\frac{\partial u_i}{\partial t} x_i + u_i \frac{\partial x_i}{\partial t}}
=  \sum_i{c_{v_i} \frac{\partial T}{\partial t} x_i + u_i \frac{\partial x_i}{\partial t}}
=  \frac{\partial T}{\partial t} \sum_i{c_{v_i} x_i + u_i \frac{\partial x_i}{\partial t}}.
\end{equation*}
Since the volume is constant---i.e., no work is done on the
system---and no heat is added, the change in the energy $U$ must be
zero. Setting $dU/dt$ to zero and solving for $dT/dt$ yields
\begin{equation*}
  \frac{dT}{dt} = - \left( \frac{1}{\sum_i{c_{v_i}x_i }}\right) \left(\sum_i
   {u_i \dot{x}_i}\right).
\end{equation*}

Note that $\frac{dT}{dt}$ is a function of both the molar
concentrations and their time derivatives. The representations of the remaining functions
$c_{v_i}(T)$ and $u_i(T)$ may be found in the literature, generally as
seven or nine term polynomials. The commonly used NASA polynomials are used in
this work~\cite{nasa}.

With the time derivative of temperature, the mathematical model of the reaction
mechanism is complete. To summarize, there is an ODE for the time derivative of each species and also
for temperature. These can be written more compactly as
\begin{equation*}
[\dot{x}_1, \dot{x}_2, \dots, \dot{x}_n,\dot{T}]^T = \mathcal{F}(\bm{x}, \dot{\bm{x}}, T)
\end{equation*}
where $\mathcal{F}$ is a nonlinear operator acting on the state
$(\bm{x}, T)$ and the time derivatives of $\bm{x}$. Note that
$\mathcal{F}$ depends on $\dot{\bm{x}}$ only through the energy
equation.

\section{Formulation of the model inadequacy}\label{sec:op}

In contrast to the simple example mechanism in \S\ref{sec:kin}, mechanisms
describing complex chemical systems like those encountered in combustion often
include hundreds of reactions.  One of the standard models for methane
combustion, for example, includes fifty-three species and 325 reactions
\cite{grimech}.  Such models are often referred to as detailed mechanisms and
will be written here as $\mathcal{D}(\bm{x}^D, \dot{\bm{x}}^D, T)$.  In the
context of a reacting flow simulation, such a large mechanism may be too
computationally expensive to be practical, and thus it is common to use a
reduced chemistry model consisting of a subset of the species and reactions from
the detailed model.

To be concrete, suppose that the detailed model includes $n_D$ species and $m_D$
reactions. Further suppose that the reduced model includes $n_R$ species and $m_R$ reactions,
where $n_R \leq n_D$ and $m_R < m_D$. The reduced model always contains fewer
reactions than the detailed; the number of species included in the reduced model
may or may not be smaller, although in practice it is almost always true that
$n_R < n_D$. The reduced model is denoted $\mathcal{R}(\bm{x}^R, \dot{\bm{x}}^R,
T)$.

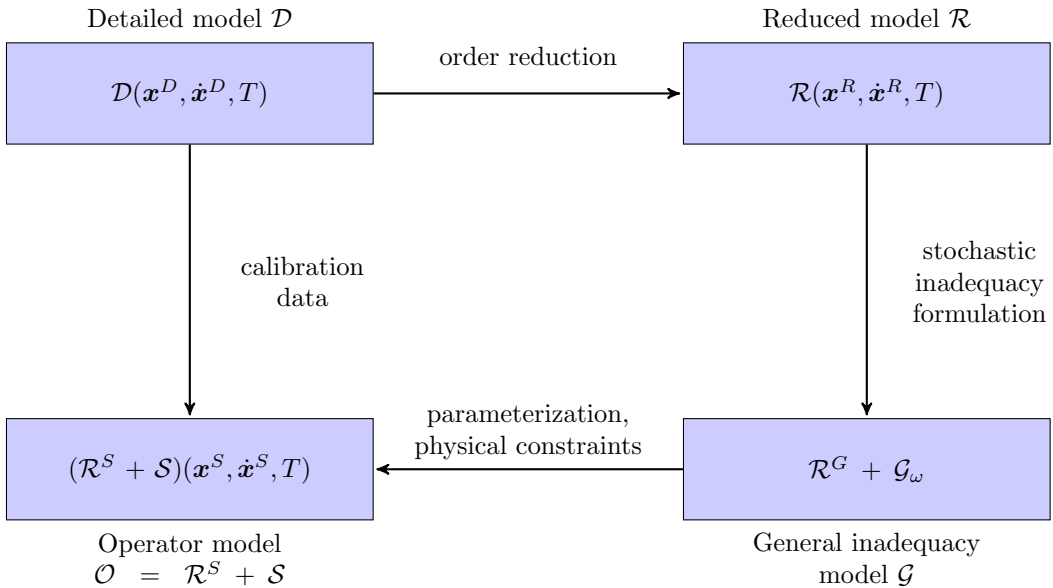
\begin{figure}[thp]
\centering
\tikzstyle{orangeblock} = [rectangle, draw, fill=orange!20, text width=5.5em, text
centered, rounded corners, minimum height=3em]%
\tikzstyle{blueblock} = [rectangle, draw, fill=blue!20, text width=12em, text
centered, minimum height=3.5em]%
\tikzstyle{clearblock} = [rectangle, text width=15em,text centered,minimum
height=3em]%
\tikzstyle{clearblocksmall} = [rectangle, text width=6em,text centered,minimum
height=3em]%
\tikzstyle{line} = [->, >=stealth',shorten >=1pt, thick]%
\tikzstyle{line2} = [<->, >=stealth',shorten >=1pt, thick]%
\begin{tikzpicture}[font=\small]%
  \node at (0,5) [blueblock] (det) {$\mathcal{D}(\bm{x}^D, \dot{\bm{x}}^D, T)$};
  \node at (9,5) [blueblock] (red) {$\mathcal{R}(\bm{x}^R,\dot{\bm{x}}^R, T)$};
  \node at (9,0) [blueblock] (ran) {$\mathcal{R}^G +
  \mathcal{G}_{\omega}$};
  \node at (0,0) [blueblock] (mat) {$(\mathcal{R}^S +
  \mathcal{S})(\bm{x}^S, \dot{\bm{x}}^S, T)$};
\draw [line] (det) -- (red);
\draw [line] (red) -- (ran);
\draw [line] (ran) -- (mat);
\draw [line] (det) -- (mat);
\node at (4.5,5.5) [clearblock] {order reduction};
\node at (10.5,2.5) [clearblock] {stochastic\\inadequacy\\formulation};
\node at (4.5,0.5) [clearblock] {parameterization,\\physical constraints};
\node at (1.5,2.5) [clearblock] {calibration\\data};

\node at (0, 6) [clearblock] {\small Detailed model $\mathcal{D}$};
\node at (9, 6) [clearblock] {\small Reduced model $\mathcal{R}$};
\node at (9, -1.2) [clearblock] {\small General inadequacy\\model $\mathcal{G}$};
\node at (0, -1.2) [clearblock] {\small Operator model\\$\mathcal{O} =
\mathcal{R}^S + \mathcal{S}$};

\end{tikzpicture}%
\caption{Relationship between the detailed model, deterministic reduced model,
  reduced model with general model inadequacy representation, and reduced model
  with stochastic operator model inadequacy representation.}
\label{fig:flow}
\end{figure}

In the case that the reduced model does not adequately represent the
detailed (or the real chemical reaction), one has two options: (1)
improve the reduced model directly with a more accurate mechanism, or (2)
incorporate a representation of the model error of the reduced model.
As noted in \S\ref{sec:intro}, it is often impractical to improve the
model and thus, the focus of this work is on developing a generally
applicable model inadequacy representation for reduced chemistry
models.  The model inadequacy is represented by a stochastic operator
$\mathcal{S}$ that is appended to the reduced model $\mathcal{R}$,
as indicated in Figure~\ref{fig:flow}, which shows the progression
from the detailed model to the proposed stochastic model.

In the figure, the $\mathcal{D}$ and $\mathcal{R}$ operators
correspond to the deterministic detailed and reduced models introduced
above.  The operator $\mathcal{R}^G$ denotes the reduced mechanism
used with a general stochastic inadequacy model,
$\mathcal{G}_{\omega}$, where $\omega$ is a set of random variables.
For example, $\mathcal{G}_{\omega}$ could be a purely statistical
model that is trained based on observations of the detailed model
reaction rates.  On the other hand, this term could represent an
augmentation of the chemical mechanism obtained by incorporating more
reactions from the detailed model.  In this case, the model inadequacy
representation would in fact be deterministic (with $\omega = \{\}$).
This approach would necessitate more information about the true
chemical reaction than we expect to have or are willing to use and
thus is not generally applicable.

Here we take an approach which is intermediate between a strictly statistical representation and a
strictly physics-based approach.  Physics is incorporated into the approach in that we insist that
the inadequacy representation respect certain known features of the system, but, since our knowledge
of the true dynamics is incomplete, the model is necessarily stochastic, implying that there is a
range of behaviors consistent with our knowledge. In the chemical kinetics case, we know that errors
in the model are due to unrepresented species and unrepresented pathways by which species transform
into each other. Further, both atoms and energy must be conserved and species concentrations and
temperature must remain non-negative. To account for the effects of the missing species and
reactions and to satisfy these constraints, a stochastic operator representation of the inadequacy
is posed, as shown on the bottom left of Figure~\ref{fig:flow}. The reduced model $\mathcal{R}^S$
plus the stochastic inadequacy operator $\mathcal{S}$ is called the operator model $\mathcal{O}$.
\revise{Note that the the physical basis of the inadequacy representation gives structure to
the operator, while the statistical aspect gives it the flexibility
needed to represent the error over a wide range of conditions and allows
it to be calibrated to data.} 

For clarity of notation, the superscripts ($D$, $R$, $G$, $S$) on either the
state vector $\bm{x}$ or the reduced model $\mathcal{R}$ are included to reflect
the model at hand.  As will become clear, to satisfy the requirements
mentioned above, the inadequacy formulation alters
the state vector $\bm{x}$ and the reduced model operator
$\mathcal{R}$, so
$\bm{x}^S$ and $\mathcal{R}^S$ differ from the corresponding $\bm{x}^R$ and
$\mathcal{R}$.

\subsection{Components and State Variables of the Operator Model}\label{sec:init}

\subsubsection{Components of the operator} 
The main action of the inadequacy operator $\mathcal{S}$ is to modify
the time derivatives of the species concentrations.  The operator
consists of three pieces: a random matrix $S$, a nonlinear operator
$\mathcal{A}$, and a nonlinear operator $\mathcal{W}$.\footnote{In
  general, a script letter refers to a nonlinear operator, capital
  letters to linear operators (matrices), lowercase bold letters to
  vectors, and lowercase (unbolded) letters to scalars.}  The random
matrix $S$ is intended to represent the most general \emph{linear}
correction that can be made to the reduced chemistry model.  However,
to ensure sufficient flexibility for mass to move between species, it
is necessary to introduce a nonlinear operator $\mathcal{A}$, as
discussed in \S\ref{sec:A}.  The final piece of the operator accounts
for conservation of energy and is denoted $\mathcal{W}$.

Note that $S$ and $\mathcal{A}$ act on just the concentrations, while $\mathcal{W}$ acts
on the concentrations and their derivatives. Moreover, it is convenient to
formulate $S$ in terms of atomic concentrations, while $\hat{S}$ denotes the
corresponding matrix in terms of molar concentrations. This section focuses on
$S$ instead of $\hat{S}$ because many of the properties
of the matrix (such as non-positivity of eigenvalues) are better expressed in terms of
atoms instead of moles (see \S\ref{app:Sprops}). To map between $S$ and $\hat{S}$, the
vector $\bm{l}$ is used, whose $i$th entry counts the number of atoms (of all
types) in one molecule of the $i$th species. For example, if the set of species is H$_2$,
O$_2$, H, O, OH, H$_2$O, then $\bm{l} = [2, 2, 1, 1, 2, 3]$.
Let $L$ denote the square matrix with the entries of $\bm{l}$ on the
diagonal.  Then $\hat{S} = L^{-1}SL$ applies to molar concentrations. Finally,
putting the three pieces together,
\begin{align*}
\mathcal{S} &= \hat{S} + \mathcal{A} + \mathcal{W} \\
            &= L^{-1} S L + \mathcal{A} + \mathcal{W},\quad \text{or more explicitly,} \\
\mathcal{S}(\bm{x}^S,\dot{\bm{x}}^S, T) &= L^{-1} S L\bm{x}^S +
\mathcal{A}(\bm{x}^S) +  \mathcal{W}(\bm{x}^S,\dot{\bm{x}}^S, T) .
\end{align*}

\subsubsection{Augmentation of the state vector} The reduced model tracks fewer
species than the detailed model. It should be possible then, for the inadequacy
formulation to represent this difference. However, we do not want to include all
the extra (missing) species in the inadequacy representation.  Therefore, in
order to account for the missing species in the reduced model, the state space
is augmented by entries for all types of atoms. These entries are referred to as
catchall species and act as a sort of pool of each atom type, representing atoms bound in
unrepresented species. The presence of
the catchall species allows the operator $\mathcal{S}$ to move atoms to and from
these pools instead of constraining every atom to one of the species of the
reduced model, which is overly restrictive because in the detailed model, atoms
may move to species that are not part of the reduced model.  Thus, $\bm{x}^S$ is
of length $n_S = n_R + n_{\alpha}$, where $n_{\alpha}$ is the number of atom
types, and is of the form \[ \bm{x}^S = [ x_1, \dots, x_{n_\alpha},x_{n_\alpha +
    1}, \dots, x_{n_\alpha + n_R }] ^T. \] 

We denote the catchall species of element $\mathbb{X}$ by
$\mathbb{X}'$.  For example, consider a reduced model that includes
H$_2$, O$_2$, OH, and H$_2$O. Then the catchall species are H$'$ and O$'$, and
\begin{equation*}
\bm{x}^S = [x_1, x_2, x_3, x_4, x_5, x_6]^T
\end{equation*}
where
$x_1,\dots,x_6$ corresponds to H$'$, O$'$, H$_2$, O$_2$, OH, and H$_2$O, in that
order.

The introduction of the catchall species raises an important point
about the structure of the reduced model: it takes on a different form
when used in conjunction with the stochastic operator $\mathcal{S}$.
Because of this fact, the reduced model used with the stochastic
operator is denoted by $\mathcal{R}^S$ to distinguish it from the
original reduced model $\mathcal{R}$.  There are two differences
between these operators.  First, $\mathcal{R}^S$ acts on a vector
space of dimension $n_S$ rather than $n_R$, although it has no effect
on the first $n_{\alpha}$ entries of $\bm{x}^S$ (i.e., the catchall
species). Second, because the effect of the catchall species on the
energy equation is not additive, the differential equation for $T$ is
removed from $\mathcal{R}^S$, and the entire calculation is accounted
for with $\mathcal{W}$.


\subsection{Physical constraints and their implications} \label{sec:constraints}
There are two non-negotiable constraints that any physically realistic
model of the system at hand must respect: (I) conservation of atoms,
and (II) non-negativity of concentrations. This ensures that the
inadequacy operator respects physical laws that are known to be true
for the systems of interest.  This section develops the implications
of these constraints for the components of the operator $\mathcal{S}$
with a focus on the matrix $S$ since, as will be shown, $\mathcal{A}$
and $\mathcal{W}$ are subsequently chosen to take forms that are known
to satisfy these constraints.

\subsubsection{Conservation of atoms}
To enforce (I), first let $E = [e_{ij}]$ be the $n_{\alpha} \times
n_S$ matrix where $e_{ij}$ is the fraction of atoms of type $i$ in one
molecule of species $j$.  Then $E S L \bm{x}^S$ is the rate of change
of the number of each atom due to the operator $S$.  Because atoms are
conserved $E S L \bm{x}^S$ must be zero, which implies that $ES = 0$.


To continue the example shown in \S\ref{sec:init}, consider the case with atom types
H and O, and species H$'$, O$'$, H$_2$, O$_2$, OH, H$_2$O. Then matrix $E$ takes the form:
\begin{equation*}
E = 
\begin{bmatrix} 
1 & 0 & 1 & 0 & 1/2 & 2/3 \\
0 & 1 & 0 & 1 & 1/2 & 1/3
\end{bmatrix}.
\end{equation*}
To satisfy the constraint that $ES = 0$, the matrix $S$ is constructed according to
\begin{equation*}
S = C P,
\end{equation*} 
where $C$ is a deterministic matrix and $P$ is probabilistic. The roll
of the matrix $C$ is to ensure conservation of atoms, while $P$ is
constructed to ensure that the concentrations are non-negative. To guarantee conservation of atoms, the
columns of $C$ must span the nullspace of $E$, i.e.  $\text{span}(C) =
\text{null}(E)$. Thus,
\begin{equation*}
ES = ECP = (EC)P = 0 \cdot P = 0.
\end{equation*}
$E$ is of dimension $n_{\alpha} \times n_S$, so the dimension of the
nullspace is $n_S - n_{\alpha} = n_R$. Thus $C$ is of dimension $n_S
\times n_R$ and $P$ is of dimension $n_R \times n_S$.

\subsubsection{Non-negativity of concentrations}
The second constraint (II) is that the concentrations must not be negative. To see
how to enforce this, consider the differential equation for species
$\mathbb{X}_i$\footnote{We drop the superscript $S$ from $\bm{x}$ here for ease of
notation.}:
\begin{equation}
\dot{x}_i
= (\mathcal{R}^S(\bm{x},\bm{T}))_i + (\mathcal{S}(\bm{x}))_i
= (\mathcal{R}^S(\bm{x},\bm{T}))_i + (L^{-1}S L\bm{x})_i  +  (\mathcal{A}(\bm{x}))_i.
\label{eq:onedx} 
\end{equation}
We must ensure that $\dot{x}_i \geq 0$ when $x_i = 0$ for $i =
1,\dots,n_S$. The first term of the RHS of (\ref{eq:onedx}) is not a
problem, as this is the nonlinear part from the reaction mechanism
and is thus already physically consistent \cite{feinberg79}.  The same
argument holds for $\mathcal{A}(\bm{x})$ because it takes the form of
a standard chemical reaction model, as will be shown in
\S\ref{sec:A}. Note that the energy operator $\mathcal{W}$ is not
written above because it does not modify the derivative of $x_i$.

Finally, the second term must satisfy the constraint. Although the
constraint is naturally posed here in terms of molar concentrations,
it is helpful to rephrase this in terms of atomic concentrations. One
can show that the constraint is satisfied for moles if and only if it
is satisfied for atoms:
\begin{equation*}
L^{-1} S L \bm{x} \geq 0 \quad \iff \quad S \bm{y} \geq 0.
\end{equation*}
To prove this, consider the $i$th entry of $L^{-1} S L \bm{x}$:
\begin{align*}
(L^{-1} S L \bm{x})_i &= L^{-1} \sum_j s_{ij}l_j x_j \\
&= \frac{1}{l_i}\sum_j s_{ij}l_j x_j
\end{align*}
but $l_j x_j = y_j$ and all $l_i > 0$, $i = 1, \dots, n_S$. Thus,
\begin{equation*}
\frac{1}{l_i}
\sum_j s_{ij}l_j x_j \geq 0 \iff \sum_j s_{ij}y_j \geq 0.
\end{equation*}
But the final term is exactly the $i$th element of $S\bm{y}$.

Continuing in terms of the atomic concentrations, the $i$th component of $S\bm{y}$ is given by
\begin{equation}
(S\bm{y})_i = s_{ii}y_i + \sum_{j \neq i} s_{ij} y_j.
\end{equation}
The first term from the diagonal, $s_{ii}y_i$, automatically respects the
constraint: $s_{ii}$ may be set to be any constant value, since then $s_{ii}y_i
\rightarrow 0$ as $y_i \rightarrow 0$. To enforce the constraint, it must be
that the sum, $\sum_{j \neq i} s_{ij} y_j$, is non-negative. But this sum
does not depend on $y_i$, so we choose to set $s_{ij} \geq 0$ for all $i \neq
j$. This could be made less restrictive by incorporating information from the
nonlinear system, i.e. set ($\mathcal{R}^S(\bm{x}))_i + \sum_{j \neq i}
\hat{s}_{ij} x_j \geq 0$, but this would violate the linearity assumption on
$S$. It would also necessitate using information from the reduced model, whereas
we aim to constrain the inadequacy operator independently of $\mathcal{R}$.

\subsubsection{Sparsity of $S$} \label{sec:sparse}
In practice, many of the entries of $S$ are
identically zero. In theory, $S$ could be completely dense if
every species included every type of atom. However, this generally does not occur in
practical combustion reactions. The following proves which
entries of $S$ are identically zero, using an argument based on the zeros of the
matrix $E$.

\begin{theorem}\label{thm:zeros}
Consider the $i$th row of $E$. Let $\mathcal{J}_i = \{j | e_{ij}  \neq 0\}$ and
$\mathcal{J}_i^c = \{j | e_{ij} = 0\}$. Then every element $s_{jk} = 0$ for $j
\in \mathcal{J}_i$ and $k \in \mathcal{J}_i^c$.
\end{theorem}
\begin{proof}
Consider the $i$th row of $E$ and the $k$th column of $S$. Since $E S = 0$, we have
\begin{equation}
0
= \sum_j e_{ij} s_{jk}
= \sum_{j \in \mathcal{J}_i} e_{ij}s_{jk} + \sum_{j \in \mathcal{J}_i^c} e_{ij} s_{jk}
= \sum_{j \in \mathcal{J}_i} e_{ij} s_{jk} + 0.
\label{eq:esum}
\end{equation}
But since $j$ and $k$ are in disjoint sets, the sum in line (\ref{eq:esum}) does not include the
diagonal term $s_{jj}$. But the diagonal term is the only negative value in the
$k$ column. Thus, all $s_{jk} = 0$, where $j \in \mathcal{J}_i$ and $k \in
\mathcal{J}_i^c$.
\end{proof}

For another method to determine the sparsity of $S$, see
Appendix~\ref{app:Sprops}.  In addition to sparsity, the constraints on $S$
imply that it has non-positive eigenvalues.  See Appendix~\ref{app:Sprops} for
more details.

\subsection{Construction of the matrix $S$}\label{sec:conS} The structure of $S$ is
now clear; the next step is to actually construct it. The challenge in this construction is that
both constraints must be simultaneously satisfied by any realization of $S$.
This subsection presents a method for construction of the operator. To help
demonstrate the upcoming matrix decompositions and inequality constraints, the
construction will also be shown for the example set of species (H$'$, O$'$,
H$_2$, O$_2$, OH, H$_2$O). In this case, $S$ has the form
\begin{equation*} 
S = 
\begin{bmatrix} s_{1,1} & 0 & s_{1,3} & 0 & s_{1,5} & s_{1,6} \\
0 & s_{2,2} & 0 & s_{2,4} & s_{2,5} & s_{2,6}\\ 
s_{3,1} & 0 & s_{3,3} & 0 & s_{3,5} & s_{3,6}\\
0 & s_{4,2} & 0 & s_{4,4} & s_{4,5} & s_{4,6}\\ 
0 & 0 & 0 & 0 & s_{5,5} & s_{5,6}\\
0 & 0 & 0 & 0 & s_{6,5} & s_{6,6}
\end{bmatrix},
\end{equation*}
where the diagonal elements are non-positive and the off-diagonal elements are
non-negative. Here, $n_R = 4$, $n_\alpha = 2$.

First, $C$ is formed to span the nullspace of $E$.  To accomplish
this, let the bottom $n_R \times n_R$ block of $C$ be the identity matrix
$I_{n_R}$. The remaining top $n_{\alpha}$ rows will be the negative of
the last $n_R$ columns of $E$. Let this matrix block be denoted
$E^*$. Note that every element of $E^*$ is non-positive. Thus, $C$ has
the form
\begin{equation*}
C = \left[ \begin{array} {c} 
E^* \\ 
\hline 
I_{n_R} 
\end{array} \right] .
\end{equation*}
Since $E = [ I_{n_{\alpha}} | - E^*]$, it is clear that $EC = 0$.

For the H$_2$/O$_2$ example at hand,
\begin{equation*} E = 
\begin{bmatrix} 1 & 0 & 1 & 0 & 1/2 & 2/3 \\
0 & 1 & 0 & 1 & 1/2 & 1/3 \end{bmatrix}. \end{equation*}
Thus,
\begin{equation*} E^* = 
\begin{bmatrix} -1 & 0 & -1/2 & -2/3 \\
0 & -1 & -1/2 & -1/3 \end{bmatrix} \end{equation*}
and
\begin{equation*} C = 
\begin{bmatrix} -1 & 0 & -1/2 & -2/3 \\
0 & -1 & -1/2 & -1/3\\
1 & 0 & 0 & 0\\
0 & 1 & 0 & 0\\
0 & 0 & 1 & 0\\
0 & 0 & 0 & 1\end{bmatrix}. \end{equation*}

Next, we construct the $n_R \times n_S$ random matrix $P$. The first step is to specify
which entries are non-negative, non-positive, or strictly zero. Then, by
taking advantage of the special structure of $C$, it is possible to transfer
the inequalities placed on the entries of $S$ to those of $P$. Let $P_1$ contain
the first $n_\alpha$ columns of $P$, and $P_2$ the remaining $n_R$ columns. So far we have
\begin{align}
S &= C P \nonumber \\
  &= \left[ \begin{array} {c}
  E^* \\ \hline
  I_{n_R} \end{array} \right]
  \left[ \begin{array} {c|c} P_1 & P_2 \end{array} \right] \nonumber \\
  &= \left[ \begin{array}{c|c}
  E^* P_1 & E^* P_2\\ \hline I_{n_R} P_1 & I_{n_R}P_2 \end{array} \right] \nonumber \\
  &= \label{eq:decomp3} \left[ \begin{array}{c}
  E^* P_1 | E^* P_2\\ \hline I_{n_R} P \end{array} \right]. 
\end{align}
The bottom row of (\ref{eq:decomp3}) shows how to transfer the inequalities from
matrix $S$ to $P$. Since $P$ is left-multiplied by the identity
matrix, it must be that the signs match for the corresponding elements of $S$.
In particular, for $ 1 < i \leq n_R$ and $\forall j$, then
\begin{align}
\label{eq:pij}p_{i,j} & \leq 0 \text { if } s_{(i+n_\alpha),j} \leq 0 \\
\label{eq:pprime}p_{i,j} & \geq 0 \text { if } s_{(i+n_\alpha),j} \geq 0 \\
\label{eq:pzero}p_{i,j} & \equiv 0 \text { if } s_{(i+n_\alpha),j} \equiv 0.
\end{align}
Thus, in the example,
\begin{equation*} P = 
\begin{bmatrix} p_{1,1} & 0 & p_{1,3} & 0 & p_{1,5} & p_{1,6} \\
0 & p_{2,2} & 0 & p_{2,4} & p_{2,5} & p_{2,6}\\ 
0 & 0 & 0 & 0 & p_{3,5} & p_{3,6}\\
0 & 0 & 0 & 0 & p_{4,5} & p_{4,6}
\end{bmatrix}, \end{equation*}
where \[p_{1,3},\, p_{2,4},\, p_{3,5},\, p_{4,6} \leq 0\] and 
 \[p_{1,1},\, p_{1,5},\, p_{1,6},\, p_{2,2},\, p_{2,5},\, p_{2,6},\, p_{3,6},\, p_{4,5}\geq 0.\] 
 Note that the number of non-zero elements in $P$ is $12$.

The three inequalities (\ref{eq:pij}-\ref{eq:pzero}) are necessary but not sufficient as this only guarantees
the inequalities of the bottom row of (\ref{eq:decomp3}) hold. The top row
introduces more restrictive inequalities on a subset of the entries of $P$.
First consider the top left block. The only nonzero elements here are the
negative entries on the diagonal. There can be no non-zero off-diagonal elements
of $S$ in this block, because each row and column correspond to a catchall
species, and atoms can never move from one catchall to another because they are
of different types, by definition. But all the entries of $E^*$ are non-positive,
and all entries of $P_1$ are non-negative by (\ref{eq:pprime}) (these correspond
to off-diagonal elements of $S$). Thus, the diagonal elements of $S$ in this top
left block are guaranteed to be non-positive, as required.

Lastly, consider the top right block: $E^* P_2$. To guarantee that these
elements are non-negative, it is necessary that the negative entries of $P_2$ (on
its diagonal) are large enough in magnitude. For these elements $s_{i,k}$ in the
top right block, $1 \leq i \leq n_\alpha$ and $n_\alpha < k \leq n_s$. Now
\begin{align*}
0 \leq s_{i,k} &= E^*_{(i,\cdot)} P_{2 (\cdot,k)}\\
        &= E^*_{(i,\cdot)} P_{(\cdot, k + n_\alpha)}\\
        &= E^*_{(i,\cdot)} P_{(\cdot, k')}\\
        &= \sum_j e^*_{i,j}p_{j,k'},
\end{align*}
where $k' = k + n_\alpha$. The only positive term above in the sum is
$e^*_{i,k}p_{k,k'}$, so this implies
\begin{equation*}
e^*_{i,k} p_{k,k'} \geq -\sum_{j \neq k} e^*_{i,j} p_{j,k'}.
\end{equation*}
A similar inequality is placed on the each element $p_{k,k'}$ for each type
of atom (each row of $E^*$ that multiplies the $k'$th column of $P$).
Therefore, to complete the set of inequalities on $P$, it is sufficient that,
for $i = 1,\dots,n_\alpha$ and $k=1,\dots,n_R$:
\begin{equation*}\label{eq:pkkmax} -p_{k,k'} \geq \frac{1}{\min_i|e^*_{i,k}|} \sum_{j\neq k}
\max_i|e^*_{i,j}| p_{j,k'}, \end{equation*}
or, in terms of the matrix $C$:
\begin{equation}\label{eq:pkkmaxC} -p_{k,k'} \geq \frac{1}{\min_i|c_{i,k}|} \sum_{j\neq k}
\max_i|c_{i,j}| p_{j,k'}. \end{equation}
For use in the following development, denote the RHS of (\ref{eq:pkkmaxC}) above as $q_{k'}$.

In the example, the extra constraints from $E^*P_2$ correspond to the diagonal
elements of $P$: $p_{1,3}$, $p_{2,4}$, $p_{3,5}$, $p_{4,6}$. For example, the
constraint $s_{1,5} \geq 0$ implies $E^*_{(1,\cdot)} P_{(\cdot,5)} \geq 0$ and  
$s_{2,5} \geq 0$ implies $E^*_{(2,\cdot)} P_{(\cdot,5)} \geq 0$. These two
constraints are then
\begin{align*}
-1 p_{1,5} &- 0 p_{2,5} - \frac{1}{2} p_{3,5} - \frac{2}{3} p_{4,5} \geq 0\\
-0 p_{1,5} &- 1 p_{2,5} - \frac{1}{2} p_{3,5} - \frac{1}{3} p_{4,5} \geq 0.
\end{align*}
Using~\eqref{eq:pkkmaxC}, the two lines above can be condensed into
the following inequality which is stronger than either:
\begin{equation*}
- p_{3,5} \geq 2 (p_{1,5} + p_{2,5} + \frac{2}{3} p_{4,5}).
\end{equation*}
Similarly, the constraints for the other negative elements take the form:
\begin{align*}
- p_{1,3} &\geq 0\\
- p_{2,4} &\geq 0\\
- p_{4,6} &\geq 3 (p_{1,6} + p_{2,6} + \frac{1}{2} p_{3,6}).
\end{align*}

\subsubsection{Transform from $P$ to $\boldmath{\xi}$}
Now each element of $P$ is of one of the following forms:
\begin{align*}
p_{i,k} &\equiv 0 \\
p_{i,k} &\geq 0 \\
-p_{i,k} &\geq q_k, \quad k = i + n_\alpha.
\end{align*}
These variables can be transformed and reindexed to a new set
$\{\xi_l\}_{l=1}^{n_\xi}$ such that the inequalities take the simple form $\xi_l
\geq 0$ for each $l$. This mapping also changes from a double-indexed system
($p_{i,j}$) to a single index ($\xi_l$). The index $l$ is introduced because the
zero elements of $P$ are not mapped to $\bm{\xi}$, so the mapping is unique to
every matrix. For
$n_\xi$ sets $\left\{l, i, k\right\}$, each $\xi_l$ is of one of the following
two forms:
  \begin{align*}
  \xi_l &= p_{i,k}, \quad k \neq i + n_\alpha\\
  \xi_l &= - (p_{i,k} + q_k), \quad k = i + n_\alpha.
  \end{align*}  
Note that the second set is of size $n_R$ and thus the size of the first set is
$n_\xi - n_R$.

For the example, $n_\xi = 12$ since there are $12$ non-zero elements of $P$.
There are $n_R = 4$ variables whose transform depends on $q_k$, and thus $n_\xi
- n_R = 8$ variables whose transform does not. The total transform is given in table~\ref{tab:xi-ex}.
\begin{table}\begin{centering} \begin{tabular}{r  l}
$\xi_i$\quad= & $p_{j,k}$\\ \hline
$\xi_1$\quad= & $p_{1,1}$\\ 
$\xi_2$\quad= & $-p_{1,3}$\\
$\xi_3$\quad= & $p_{1,5}$\\
$\xi_4$\quad= & $p_{1,6}$\\
$\xi_5$\quad= & $p_{2,2}$\\
$\xi_6$\quad= & $-p_{2,4}$\\
$\xi_7$\quad= & $p_{2,5}$\\
$\xi_8$\quad= & $p_{2,6}$\\
$\xi_{9}$\quad= & $-p_{3,5} - \frac{2}{3}(p_{1,5} + p_{3,5} + \frac{2}{3}p_{4,5})$\\ 
$\xi_{10}$\quad= & $p_{3,6}$\\ 
$\xi_{11}$\quad= & $p_{4,5}$\\ 
$\xi_{12}$\quad= & $-p_{4,6} - 3 (p_{1,6} + p_{2,6} + \frac{1}{2}p_{3,6})$\\ 
\end{tabular}
\caption{The transformed variables $\bm{\xi}$ for the example operator. \label{tab:xi-ex}}
\end{centering}
\end{table}

To complete the construction, it remains to specify the probability distribution that
governs each variable $\xi_l$. Since $\xi_l \geq 0, l=1,\dots,n_\xi$, let
\begin{equation} \label{eq:xi-dist}
\xi_l \sim \log\mathcal{N}(\mu_l^\xi, \eta_l^\xi).
\end{equation}
The role of the hyperparameters $\mu$ and $\eta$ and how to calibrate
them will be explained in detail in \S\ref{sec:cal}. For ease and
generality of notation, let $\bm{\psi}$ be the vector of inadequacy
parameters (so far, $\bm{\psi} = \bm{\xi}$ but more inadequacy
parameters will be introduced in the upcoming subsections), and let
$\bm{\zeta}$ be the vector of all hyperparameters.


This concludes the description of $S$. Recall that the operator consists of
three pieces:
\begin{equation}
\mathcal{S} = \hat{S} + \mathcal{A} + \mathcal{W}.
\end{equation}
The next subsections continue with formulations of $\mathcal{A}$ and $\mathcal{W}$.

\subsection{The catchall reactions $\mathcal{A}$} \label{sec:A}
There is much flexibility in the matrix $S$ with respect to how it can
redistribute atoms from certain concentrations to others. In fact, it is the
most flexible (or general) linear formulation. That is, at every point in time,
a certain species $\mathbb{X}_i$ can be redistributed among all other species
$\mathbb{X}_j$ as long as $s_{ji} > 0$. 
Moreover, the rates at which these processes occur are not set a priori, but are
calibrated using the available data. The random matrix
$S$ also provides the flexibility of the catchall species---allowing a place
for atoms to go that might in fact make up a species not included in
$\mathcal{R}$ but present in $\mathcal{D}$.

However, there is one serious limitation of $S$ due entirely to the linearity:
while any species can move to the catchall species (e.g.,
\cee{H$_2$O  -> 2H$'$ + O$'$}), a catchall species can only directly move to a species made up of
the same type of atom. Therefore, a reaction like the reverse of the previous,
namely \cee{ 2H$'$ + O$'$ -> H$_2$O },
is not allowed. This would require a term that depends on the concentrations of
both catchall species, but in a linear operator this is not possible.  In some cases, this limitation is not serious.  For example, in the case with species H$_2$, O$_2$, OH, and H$_2$O, the catchall species could
move back to the reduced set of species since H$'$ could form H$_2$ and
O$'$ could form O$_2$.

However, this movement from catchall species back to real species is not always possible.  Consider a methane combustion model that includes the species H$_2$, O$_2$, H$_2$O, CH$_4$, CO,
and CO$_2$. Then the operator model species set is H$'$,
O$'$, C$'$, H$_2$, O$_2$, H$_2$O, CH$_4$, CO, and CO$_2$. Here, $S$ can send
carbon atoms from CH$_4$, CO, and CO$_2$ into C$'$. But then they are stuck because
C$_n$, for any $n=1,2,\dots$, is not in the reduced set of species. To overcome the
linearity limitation, we introduce a straightforward but nonlinear modification to the
operator: for any species $\mathbb{X}_i$ that is made up of more than one type of
atom, a nonlinear reaction is included in which the product is $\mathbb{X}_i$
and the reactants are the corresponding catchall species. For example, in the methane case,
\begin{align*}
\cee{ 2H$'$ + O$'$ &->[\kappa_1] H$_2$O }\\
\cee{ 4H$'$ + C$'$ &->[\kappa_2] CH$_4$ }\\
\cee{ O$'$ + C$'$ &->[\kappa_3] CO }\\
\cee{ 2O$'$ + C$'$ &->[\kappa_4] CO$_2$ }.
\end{align*}
This set of reactions is represented by the nonlinear operator $\mathcal{A}$.
Note that the form is analogous to a general reaction model. Thus, the
constraints (I) and (II) are automatically satisfied.

This modification introduces $n_\kappa$ reaction rate coefficients $\kappa$ to be
calibrated. Similar to the variables $\bm{\xi}$, each $\kappa$ is positive, by
design. Thus,
\begin{equation} \label{eq:kappa-dist}
\kappa \sim \log\mathcal{N}(\mu^{\kappa}, \eta^{\kappa}).
\end{equation}
Then $\bm{\psi}$ is augmented to include these rate coefficients $\kappa$, and
$\bm{\zeta}$ is augmented to include the additional hyperparameters $\mu^{\kappa}$
and $\eta^{\kappa}$.

\subsection{The energy operator $\mathcal{W}$}
The third and final component of the operator is the nonlinear stochastic energy
operator $\mathcal{W}$. The role of $\mathcal{W}$ is to account for temperature changes due to
atoms moving into and out of the catchall species. In other words, allowing for
the existence of the catchall species endows them with mass; here 
the catchall formulation is completed by endowing them with internal energy.

Recall the differential equation for $dT/dt$:
\begin{equation*} \frac{dT}{dt} = \mathcal{W}(\bm{x},\dot{\bm{x}}, T) = -\left(
\frac{1}{\sum_i^{n_S}{c_{v_i}(T) x_i }}\right) \left(
\sum_i^{n_S}{u_i(T) \dot{x}_i} \right). \end{equation*}
For $n_\alpha < i \leq n_S$, $c_{v_i}(T)$ and $u_i(T)$ are known as functions of
temperature from the literature on thermodynamic properties of chemical species
\cite{nasa}. The new contribution is to allow for $u_i(T)$ and  $c_{v_i}(T)$ for
$i=1,\dots,n_\alpha$, that is, allow for catchall energies and specific heats
and then incorporate these into the calculation of the time derivative of
temperature. For actual chemical species, these properties are always given as a
function of temperature. Thus, each new coefficient will also be allowed to
have a simple temperature-dependence. Consider a catchall species
$\mathbb{X}'_i$, $i=1,\dots,n_\alpha$. For the internal energy, we pose the
following form:
\begin{align*}
  u_i(T) &= \alpha_{0_i} + \alpha_{1_i} T + \alpha_{2_i} T^2,
\end{align*}
and, since $c_v$ is its derivative with respect to temperature,
\begin{equation*}
  c_{v_i}(T) = \alpha_{1_i} + 2\alpha_{2_i} T.
\end{equation*}
Then $\alpha_0$, $\alpha_1$, and $\alpha_2$ are additional parameters to be
calibrated. Furthermore, like all the other random variables introduced during
the modeling of the inadequacy operator, each will in fact be represented by a
probability distribution. This is appropriate since we have incorporated some
physical information (temperature-dependence), but the true functional form is
uncertain. It is known that $\alpha_1$ and $\alpha_2$ are
positive, while $\alpha_0$ could be positive or negative. These properties are
exhibited in probability densities of the form
\begin{align}
  \alpha_0 &\sim \mathcal{N}(\mu^\alpha_0,\eta^\alpha_0)  \label{eq:alpha0-dist}\\
  \alpha_l &\sim \log\mathcal{N}(\mu^\alpha_l,\eta^\alpha_l), \quad l = 1, 2.
  \label{eq:alphal-dist}
\end{align}
Since the above applies to the $n_\alpha$ catchall species, there are $3 n_\alpha$
new variables. Of course, $\bm{\psi}$ and
$\bm{\zeta}$ are again augmented to include the new (and final) inadequacy
parameters and hyperparameters. Thus, the inadequacy parameters are $\bm{\psi} = \{\bm{\xi},
\bm{\kappa}, \bm{\alpha}\}$, and the hyperparameters are $\bm{\zeta} =
\{\bm{\mu^\xi}, \bm{\eta^\xi}, \bm{\mu^\kappa}, \bm{\eta^\kappa},
\bm{\mu^\alpha}, \bm{\eta^\alpha}$\}.

This concludes the description of the stochastic operator $\mathcal{S}$. Many
model parameters and hyperparameters have been introduced for the formulation of
the reduced and stochastic operator models; the calibration of these parameters
and validation of the models is discussed in \S\ref{sec:cal}.

\subsection{Mapping from the operator to typical reaction form} 
For given values of the inadequacy parameters $\bm{\psi}$, the operator as
constructed may be interpreted as providing a new, enriched chemical
reaction model relative to the original reduced model.  This fact is
important because it allows relatively straightforward implementation
of the stochastic operator model inadequacy approach in existing
software for solving chemical systems.  Further, it provides an avenue
for physical interpretation of the operator and associated calibration
results.

Since the nonlinear operators $\mathcal{A}$ and $\mathcal{W}$ are
constructed in the usual fashion in this modeling domain, there is
nothing to show.  However, the interpretation of the action of the
operator $\hat{S}$ as a set of chemical reactions may not be
immediately clear.  It is now demonstrated that the random matrix
$\hat{S} = L^{-1}SL$ can be mapped to a typical chemical reaction of
the form $\cee{A ->[k] \sum{ \beta B}}$.

\begin{theorem}\label{thm:op-map}For every $j=1,\dots,n_S$, the $j$th column of $\hat{S}$ corresponds
to the reaction 
\begin{equation}
\cee{X$_j$ ->[k_j] $\sum_{p \neq j}{\beta_{jp}}$ X$_p$},
\end{equation}
where $k_j = \abs{\hat{s}_{jj}}$ and
$\beta_{jp} = \frac{\hat{s}_{jp}}{\abs{\hat{s}_{jj}}}$.
\end{theorem}
\begin{proof}
  Let $\bm{x}$ be the vector of concentrations of length $n$ (drop the subscript
  $S$ for ease of notation). Let the set of reactions above be denoted
  $\mathcal{L}(\bm{x})$ (in the same way that the reduced mechanism model is written
  $\mathcal{R}(\bm{x})$). We will show $\hat{S}\bm{x} = \mathcal{L}(\bm{x})$,
  element-wise.

  First,
  \begin{equation}
    \hat{S}\bm{x}  
    = \begin{pmatrix} \hat{s}_{1,1}x_1 + \hat{s}_{1,2}x_2 + \dots +
      \hat{s}_{1,n}x_n\\
      \hat{s}_{2,1}x_1 + \hat{s}_{2,2}x_2 + \dots +
      \hat{s}_{2,n}x_n\\ \vdots \\
      \hat{s}_{n,1}x_1 + \hat{s}_{n,2}x_2 + \dots +
    \hat{s}_{n,n}x_n \end{pmatrix},
  \end{equation}
  and for a single species $\mathbb{X}_i$,
  \begin{equation}
    (\hat{S}\bm{x})_i = \sum_j \hat{s}_{ij}x_j.
  \end{equation}

  Now consider $\mathcal{L}(\bm{x})$. The rate for a particular
  $\mathbb{X}_i$ consists of multiple terms: one in which $\mathbb{X}_i$ is the
  chemical reactant, and $n-1$ terms in which $\mathbb{X}_i$ is the chemical
  product. When $\mathbb{X}_i$ is a reactant, the corresponding rate is $-k_i
  x_i = \hat{s}_{ii}x_i$. When $\mathbb{X}_i$ is a product (and $\mathbb{X}_j$
  is the reactant, $j \neq i$), the rates from each reaction are
  \begin{equation}
    + k_j \beta_{ij} x_j = \abs{\hat{s}_{jj}} \left(
    \frac{\hat{s}_{ij}}{\abs{\hat{s}_{jj}}}\right) x_j = \hat{s}_{ij}x_j, \quad j\neq i.
    \end{equation}
    Putting the two terms together, we have
    \begin{align}
      (\mathcal{L}(\bm{x}))_i &= \hat{s}_{ii}x_i + \sum_{j \neq i} \hat{s}_{ij} x_j \\
                &= \sum_j \hat{s}_{ij} x_j \\
                &= (\hat{S}\bm{x})_i.
    \end{align}
\end{proof}

\section{Calibration and validation}\label{sec:cal}
This section describes a Bayesian approach to model calibration and
validation for the reduced model with the stochastic inadequacy
operator described in \S~\ref{sec:op}.  Since our focus is
model inadequacy, we take the reduced model, including rate parameter
values, as originally specified, meaning that the rate parameters in
the reduced model are not part of the calibration procedure.
Certainly, one could choose to infer the rate parameters and model
inadequacy simultaneously, but this inference is beyond the scope of
the current paper.  Instead, we focus on inferring the hyperparameters
of the inadequacy operator.  This inference can be accomplished in a
straightforward manner using a hierarchical Bayesian approach.  This
approach, including details of the data used and prior and likelihood
forms, is described in \S~\ref{sec:calS}.  Techniques from posterior
predictive model assessment, which are used to (in)validate both
the original reduced model and the reduced model enriched with the
inadequacy operator, are discussed in \S~\ref{sec:cal-val}.

\subsection{Calibration of the inadequacy operator model}
\label{sec:calS}
As shown in \S~\ref{sec:op}, the parameters of $S$, $\mathcal{A}$, and $\mathcal{W}$ are
characterized by probability distributions whose hyperparameters must be
calibrated.  Because the primary parameters
of interest are actually hyperparameters characterizing the probability density
associated with the parameters that appear directly in the model, it is natural
to pose the calibration problem within the hierarchical Bayesian modeling
framework described in the work of Berliner \cite{berliner96, wikle01}.

\subsubsection{Hierarchical Bayesian modeling}
The calibration problem is formulated as a single Bayesian update for
the hyperparameters $\bm{\zeta}$ and the inadequacy model parameters
$\bm{\psi}$, given the observations $\bd$.  Bayes' theorem requires
that
\begin{equation*}
p(\bm{\psi}, \bm{\zeta} | \bd)
\propto
p(\bd | \bm{\psi}, \bm{\zeta}) \, p(\bm{\psi}, \bm{\zeta}).
\end{equation*}
This form can be simplified using the hierarchical structure of the
model.  First, the map from $(\bm{\psi}, \bm{\zeta})$ to the
observables does not depend directly on $\bm{\zeta}$, the values of
$\bm{\zeta}$ are irrelevant after conditioning on $\bm{\psi}$.  Thus,
the likelihood becomes
\begin{equation*}
p(\bd | \bm{\psi}, \bm{\zeta}) = p(\bd | \bm{\psi}).
\end{equation*}
Second, because the distribution for $\bm{\psi}$ depends on $\bm{\zeta}$, it is
convenient to rewrite the joint prior as
\begin{equation*}
p(\bm{\psi}, \bm{\zeta})
= p(\bm{\psi} | \bm{\zeta}) \, p(\bm{\zeta}).
\end{equation*}
Thus, the posterior distribution can be written as
\begin{equation*}
p(\bm{\psi}, \bm{\zeta} | \bd) 
\propto
p(\bd | \bm{\psi}) \, p(\bm{\psi} | \bm{\zeta}) \, p(\bm{\zeta}).
\end{equation*}
Clearly, the posterior represents the joint distribution for the
hyperparameters $\bm{\zeta}$ and the inadequacy parameters $\bm{\psi}$
conditioned on the data.  However, the particular values of the
inadequacy parameters $\bm{\psi}$ that are preferred by the given data
are not necessarily of interest because the goal is for the
formulation to be applicable to a broad range of problems, including
scenarios outside the calibration data set.  In this situation, the
hyperparameters $\bm{\zeta}$ are the primary target of the inference
rather than $\bm{\psi}$, and one can marginalize over $\bm{\psi}$ to
find the joint posterior distribution for hyperparameters:
\begin{equation*}
p(\bm{\zeta} | \bd) = \int p(\bm{\psi},\bm{\zeta} | \bd) d\bm{\psi}.
\end{equation*}
This joint posterior is equivalent to that found by formulating the following
inverse problem:
\begin{equation*}
p(\bm{\zeta} | \bd) \propto p(\bd | \bm{\zeta}) \, p(\bm{\zeta}),
\end{equation*}
where the likelihood is given by
\begin{equation*}
p(\bd | \bm{\zeta})
=
\int p(\bd | \bm{\zeta}, \bm{\psi}) p(\bm{\psi} | \bm{\zeta}) d\bm{\psi}.
\end{equation*}

\subsubsection{Inverse problem details}
The data that is to be used in the Bayesian update described above are
observations generated by the detailed chemical kinetics model.  Specifically,
the calibration data set consists of observations of the molar concentrations of
each of the $n_R$ species tracked by the reduced model and temperature, at $n_t$
instances in time, and for $n_{IC}$ initial conditions.  The initial condition
is given by the set $\{x_f, x_o, T\}|_{t=0}$ and can be characterized by just
two quantities: the equivalence ratio $\phi$ and initial temperature $T_0$.  The equivalence ratio
quantifies how far the initial condition deviates from the stoichiometric ratio
of fuel to oxidizer and is defined by
\begin{equation*}
\phi = \frac{x_f/x_o}{x_{f_{STO}}/x_{o_{STO}}},
\end{equation*}
where $x_{f_{STO}}$ and $x_{o_{STO}}$ denote the stoichiometric concentrations
of fuel and oxidizer, respectively.  Thus, the initial condition is written as
the set $IC = \{\phi, T_0\}$.

The true value of the observable (i.e., the output of the detailed
model) is denoted by $d^t$ and may be indexed as follows:
\begin{align*}
  d^t_{ijl} &= x_i^D(t_j,IC_l), \quad i=1,\dots,n_R; \quad j=1,\dots,n_t; \quad
    l=1,\dots,n_{IC};\\
    d^t_{ijl} &= T^D(t_j,IC_l), \quad i=n_R+1.
\end{align*}
For the purposes of calibration, the observations are collected into
the vector $\bd$, and it is assumed that data are contaminated by
additive Gaussian noise, such that
\begin{equation*}
\bd = \{d_{ijl}\},
\end{equation*}
where
\begin{equation*}
d_{ijl} = d^t_{ijl} + \epsilon_{ijl}.
\end{equation*}
and $\epsilon_{ijl} \sim \mathcal{N}(0,\sigma_{\epsilon_{ijl}}^2)$, where $\sigma_{\epsilon_{ijl}} = \sqrt{0.001} \approx
0.032$ if $i=1,\dots,n_R$ and $\sigma_{\epsilon_{ijl}} = \sqrt{1000} \approx 32$ if $i=n_R+1$.
For simplicity
of exposition in the following, this can be reindexed:
\begin{equation*} \label{eq:datamod}
d_i = d^t_i + \epsilon_i, \quad i = 1, \dots, n_d,
\end{equation*} 
where $n_d = (n_R + 1) n_t n_{IC}$.

To formulate the likelihood function, consider the mapping
$\mathcal{M}^S$ from the model inadequacy parameters $\bm{\psi}$ to
the observables that is induced by the reduced chemistry model after being
enriched by the stochastic operator inadequacy representation.
The model claims that
\begin{equation*}
d_i = \mathcal{M}^S_i(\bm{\psi}) + \epsilon_i, \quad i = 1, \dots, n_d.
\end{equation*}
Thus, the likelihood is
given by
\begin{align*}
p(\bd| \bm{\psi},\bm{\zeta}) &=p(\bd
|\bm{\psi}) \\
&= \frac{1}{(2 \pi)^{n_d/2} |\Sigma|^{1/2}}
\exp\left\{-\frac{1}{2}(\bd - \mathcal{M}^S(\bm{\psi}) )^T\Sigma^{-1}(\bd
- \mathcal{M}^S(\bm{\psi}))\right\},
\end{align*}
where $\Sigma$ is the covariance matrix for $\epsilon_i$.

Following the hierarchical scheme, the joint prior for the inadequacy
parameters and hyperparameters is given by
\begin{equation*}
    p(\bm{\psi}, \bm{\zeta}) = p(\bm{\psi} | \bm{\zeta}) \, p(\bm{\zeta}).
\end{equation*}
The conditional prior distribution $p(\bm{\psi}|\bm{\zeta})$ for
the inadequacy parameters given the hyperparameters is implied by the
proposed structure in \S~\ref{sec:op}, given in lines
(\ref{eq:xi-dist}), (\ref{eq:kappa-dist}), (\ref{eq:alpha0-dist}),
(\ref{eq:alphal-dist}). \revise{For example, several of the inadequacy parameters are required to
be non-negative to satisfy the physical constraints; these are modeled as log-normal distributions.
The few unconstrained inadequacy parameters are modeled as normal distributions.} Thus, the
conditional prior distribution of each inadequacy parameter is the following:
\begin{align*}
  \xi_i &\sim \log \mathcal{N}(\mu^\xi_i,\eta^\xi_i), \quad i=1,\dots,n_\xi\\
  \kappa_i &\sim \log \mathcal{N}(\mu^\kappa_i,\eta^\kappa_i), \quad
  i=1,\dots,n_\kappa\\
  \alpha_{0_i} &\sim \mathcal{N}(\mu^\alpha_{0_i},\eta^\alpha_{0_i}), \quad
  i=1,\dots,n_\alpha\\
  \alpha_{l_i} &\sim \log \mathcal{N}(\mu^\alpha_{l_i},\eta^\alpha_{l_i}), \quad
l=1,2,\,\, i=1,\dots,2n_\alpha.
\end{align*}
Recall that $\bm{\xi}$ are the inadequacy parameters of $S$, $\bm{\kappa}$ are those
of $\mathcal{A}$, and $\bm{\alpha}$ of $\mathcal{W}$.

The hyperparameters are taken to be independent in the prior with the
following prior distributions:
\begin{align*}
  \mu^{(\cdot)}_i &\sim \mathcal{N}(\mu^{\mu^{(\cdot)}}_i, \eta^{\mu^{(\cdot)}}_i)\\
\eta^{(\cdot)}_i &\sim \mathcal{J}(0,\infty),
\end{align*}
where $(\cdot)$ represents $\xi$, $\kappa$, or $\alpha$, and
$\mathcal{J}$ denotes the Jeffreys distribution $p_{\mathcal{J}}(x)
\sim \mathcal{J}(0,\infty)$, which is given by
\begin{equation*}
p_{\mathcal{J}}(x) = \frac{1}{x}, \quad x \in (0, \infty).
\end{equation*}
Although the Jeffreys distribution is not normalizable, it is still a
valid prior \cite{degroot11}. \revise{Correlations between inadequacy parameters are not represented in
the prior distributions because we have no {\it a priori} information
about them.
However, correlations may be discovered through the inference process. Understanding the inferred
correlation structure from a chemistry perspective would be an interesting avenue of research, but
is out of scope for this work.}


The posterior distribution $p(\bm{\psi},\bm{\zeta} | \bd)$ can be
sampled using Markov chain Monte Carlo sampling methods
\cite{cowles96,gilks96,haario06}.  The algorithm used in this work is
the Delayed Rejection Adaptive Metropolis (DRAM) algorithm
\cite{haario06}.  Specifically, the results shown in \S~\ref{sec:ex}
were generated using the DRAM implementation in the QUESO
(Quantification of Uncertainty for Estimation, Simulation, and
Optimization) library\cite{estacio13, prudencio12}.  QUESO is designed
to enable research in Bayesian statistics by providing sampling
algorithm implementations that can be used in parallel computing
environments.
%
There are other software libraries available to sample
posterior distributions including BUGS (Bayesian Inference Using Gibbs
Sampling) \cite{lunn00} and MUQ (MIT Uncertainty Quantification
library) \cite{muq}.

\subsection{Validation}\label{sec:cal-val}
Once a model has been constructed and calibrated, the next step is to
validate or assess the consistency between observations of the modeled
system and the calibrated model. The validation approach used here is
that of posterior predictive assessment \cite{gelman96, rubin98}.

Consider a set of observations of the system $\{v_i\}_{i=1}^{n_v}$. This set will in
general include the data $\bd$ used for calibration and may also
include additional observations of the same or different quantities
not used in calibration. However, there is an observational error
$\epsilon$ for each observation so that the observed value $v_i$ is
related to the unknown true value $v^t_i$ by
\begin{equation*}
v_i=v_i^t+\epsilon_i.
\end{equation*}
The relevant validation question is whether the
observations $v_i$ are consistent with the model's claim regarding the
observation.

Denoting the model parameters as $\bm{\theta}$, the prediction of the
observed value $v_i$ according to the calibrated model is
%
\begin{equation}\label{eq:post-v}
p(v_i|{\bd})=\int_{d_t}p_\epsilon(v_i-v_i^t)\left(\int_{\bm{\theta}}p(v^t_i|{\bm{\theta}})p({\bm{\theta}}|{\bd})\,d
{\bm{\theta}}\right)dv^t_i .
\end{equation}
Here, depending on the circumstance, the parameters ${\bm{\theta}}$
may include physical parameters as well as inadequacy parameters
$\bm{\psi}$ and hyperparameters $\bm{\zeta}$.

Thus, the validation question is whether the available observations
are consistent with the model distribution $p(v_i|{\bd})$.  In the
case of the chemical kinetics models considered here, two different
validation situations are relevant. First, in testing the reduced
model itself (without inadequacy), $\bm{\theta}$ includes only the
kinetic parameters $\bm{k}$.  If these parameters are treated as
deterministic and known, then $p(\bm{\theta} | \bd)$
in~\eqref{eq:post-v} is the delta distribution, and the only
uncertainty about $v_i$ comes from observational error.  This approach
is used in \S~\ref{sec:ex_step23} to assess the reduced model alone.


Second, in developing a predictive model, one must assess whether the
stochastic inadequacy representation can account for model
discrepancies over a broad range of conditions, particularly for
conditions not included in the calibration.  In this situation, the
question is not whether there exists a $\bm{\psi}$ that can correct
the model for a given scenario but rather whether the uncertainty
in $\bm{\psi}$ induced by the stochastic form parameterized by
$\bm{\zeta}$ is sufficient to characterize the mismatch between the
data and reduced model for the whole range of scenarios of interest.
In this case then, $\bm{\zeta}$ is the target of the calibration
problem and the posterior for $\bm{\psi}$ is not used.  Thus, the
distribution for $v_i^t$ is given by
\begin{equation*}
p(v_i^t|{\bd})=\int_{\bm{k}} \int_{\bm{\psi}}
\int_{\bm{\zeta}} p(v_i^t | \bm{k, \psi, \zeta})
p(\bm{\psi} | \bm{\zeta}) p(\bm{k, \zeta}|\bd) d\bm{\zeta} d\bm{\psi}
d\bm{k}.
\end{equation*}
This approach is used in \S~\ref{sec:ex_step56}.

In each of these cases, the integral (\ref{eq:post-v}) yields the
posterior prediction of the observation $v_i$ which can be used to
find the total probability of observing a value less probable than the
actual observation. As explained in \cite{Oliver_2015}, this
probability can be used as a validation metric, which in turn makes
use of highest probability density (HPD) credibility regions
\cite{box73}. The $\beta$-HPD ($0 \leq \beta \leq 1$) credibility
region $\mathbb{S}$ is the set for which the probability of belonging
to $\mathbb{S}$ is $\beta$ and the probability density for any point
inside $\mathbb{S}$ is higher than those outside. Define for one
observation $v_i$,
\begin{equation*}
\gamma_i = 1 - \beta_{\text{min}_i},
\end{equation*}
where
$\beta_{\text{min}_i}$ is the smallest value of $\beta$ for which $v_i \in
\mathbb{S}_i$. Another way to think of $\gamma_i$ is that it is the integral of
$p(v_i^t | \bd)$ over the domain $\mathcal{V}_i = \{v_i^t: p(v_i^t | \bd) <
p(v_i| \bd)\}$. For samples 
$\{\bm{v}_{ij}\}_{j=1}^J$ of this distribution $p(v_i^t | \bd)$, we have 
\begin{equation*}
\gamma_i 
=
\int_{\mathcal{V}_i} p(v_i^t | \bd) dv_i^t
\approx
\frac{1}{J}\sum_j 1_{\bm{v}_{ij} \in \mathcal{V}_i}.
\end{equation*}
A delicate point here is the choice of tolerance level $\tau$ for
which, if $\gamma < \tau$, the model is deemed inconsistent with the
observation(s). A typical value for the tolerance is 0.05, although
there is an extensive discussion in the statistics literature about
how to interpret this \cite{gelman04, miller81, Oliver_2015}. When
comparing multiple observations but treating them as independent, as
we will later on, the tolerance should be corrected and set lower
because with many observations of a random variable it is more likely
to make a low-probability observation. The Bonferroni correction
suggests dividing the tolerance by the number of points
\cite{bonferroni35}. Ideally, all data points will be clearly
consistent with the model output (the model is not invalidated), or,
at least one will be clearly inconsistent (the model is invalid and
thus inadequate).

\section{Hydrogen combustion} \label{sec:ex} As an example,
the proposed inadequacy operator approach is applied to a chemical mechanism model of hydrogen
combustion. Since there are several stages of the process, it is helpful to summarize
the steps:
\begin{enumerate}
\item Identify the reduced kinetics model and a data source, which can be a detailed model if one exists (\S \ref{sec:ex_step1}).
\item Use a predictive assessment to (in)validate the reduced model (\S \ref{sec:ex_step23}).
\item If invalid, represent the inadequacy using the stochastic operator method (\S \ref{sec:form}).
\item Calibrate the parameters of the stochastic operator using data (\S \ref{sec:ex_step56}).
\item Use a posterior predictive check to validate the new model (\S \ref{sec:ex_step56}).
\item Make a prediction (if not invalidated) (\S \ref{sec:ex_step7}).
\end{enumerate}

\subsection{Identification of the reduced model and data source} \label{sec:ex_step1}
We investigate a reduced model of H$_2$/O$_2$ combustion proposed by
Williams~\cite{williams08}.  For purposes of illustration, data are
generated according to a detailed model also proposed by
Williams~\cite{williams08}. In the detailed model, there are two types of
atoms: hydrogen and oxygen; eight distinct species: H$_2$, O$_2$, H,
O, OH, HO$_2$, H$_2$O, H$_2$O$_2$; and twenty-one elementary
reactions. The reduced model contains only seven of these species and
five reactions. The resulting differential equations are much simpler than those given
by the full model. Both the twenty-one and
five reaction mechanisms and corresponding forward reaction rate parameters are
listed in appendix \ref{app:mech}. 

\subsection{Invalidation of the reduced model} \label{sec:ex_step23}
Since we choose to view the reduced model as including both the form
(i.e., chosen species and reactions) and parameter values given by
the original authors, the reduced model is not calibrated.  Instead,
the rate parameters are taken to be deterministic with the values
assigned by~\cite{williams08}.  Thus to assess the validity of the
reduced model, one must simply compare its predictions of the
observations, which are based on the deterministic model predictions
and the stochastic description of the observation error, with data.
For purposes of this comparison, observations are generated using the
detailed model for each of the seven species tracked by the reduced
model ($n_R = 7$) plus temperature, at five instances in time ($n_t =
5$), and for one initial condition ($n_{IC} = 1$).
Thus, $n_d = 8
\times 5 \times 1 = 40$.  The set of time points is $\{20, 40, 60,
80,100\}\SI{}{\micro \second}$, and the initial condition is $\phi =
1.0$, $T_0 = \SI{1300}{\kelvin}$.
  The observational
error is modeled as Gaussian with standard deviation $\sigma =
\sqrt{0.001}$ mol/m$^3$ for the species concentrations and $\sigma = \sqrt{1000}$ K for
the temperature.

\begin{figure}
  \centering
  \begin{subfigure}{.49\textwidth}
    \centering
    \includegraphics[width=\textwidth]{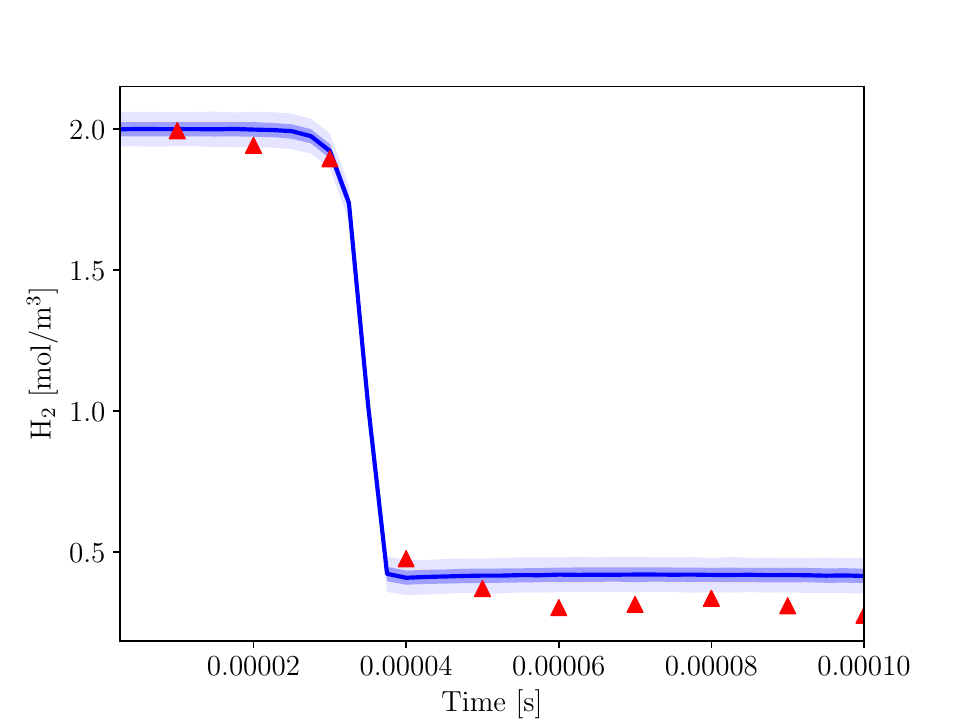}
  \end{subfigure}
  \begin{subfigure}{.49\textwidth}
    \centering
    \includegraphics[width=\textwidth]{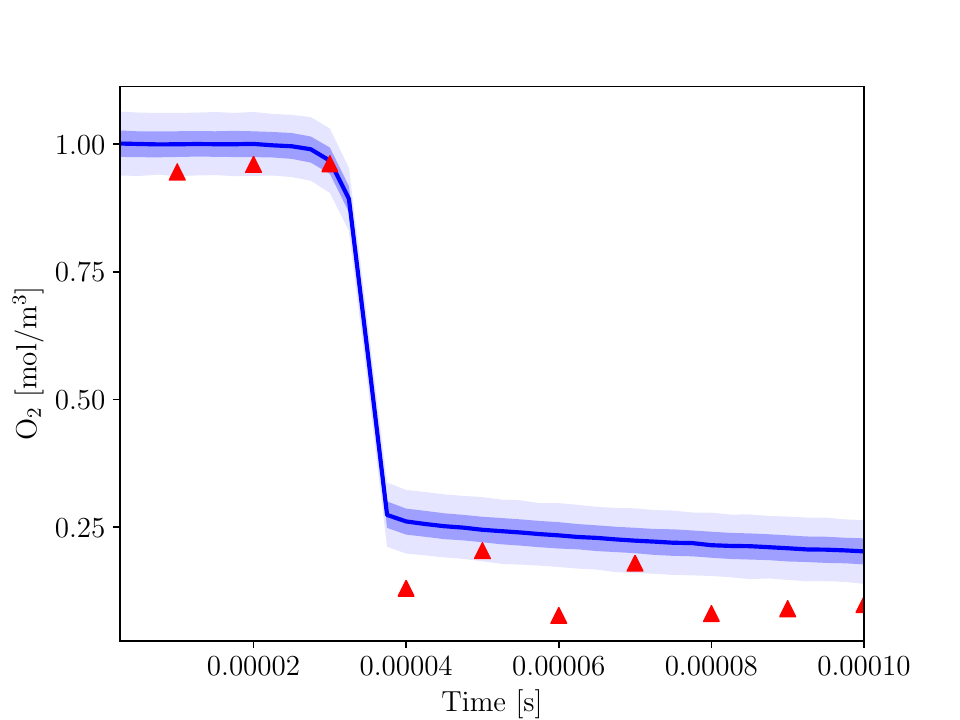}
  \end{subfigure}\\
  \begin{subfigure}{.49\textwidth}
    \centering
    \includegraphics[width=\textwidth]{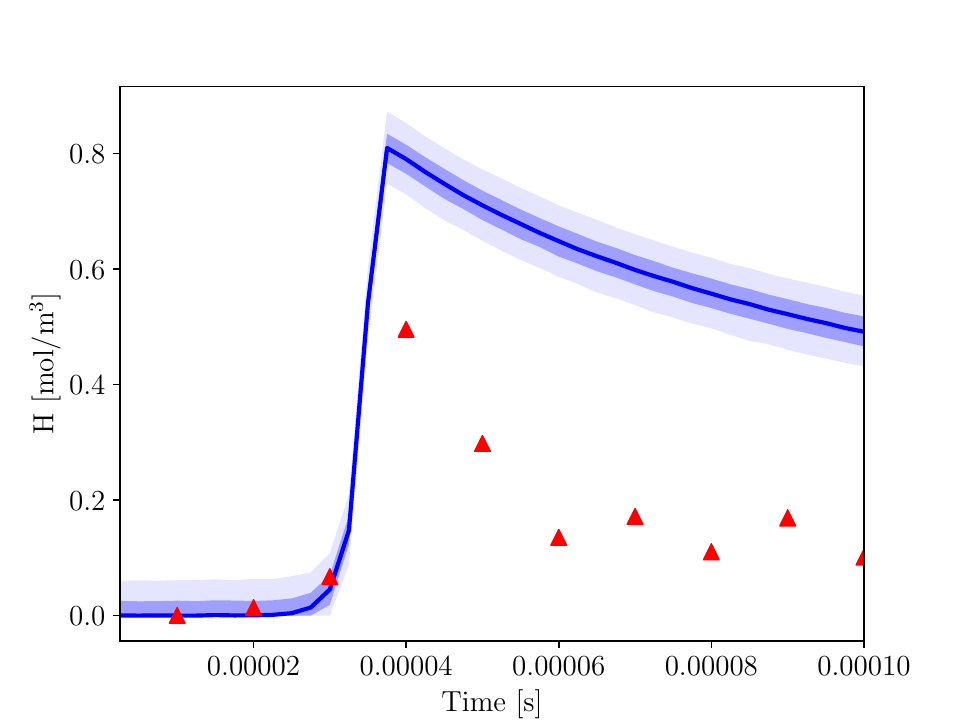}
  \end{subfigure}
  \begin{subfigure}{.49\textwidth}
    \centering
    \includegraphics[width=\textwidth]{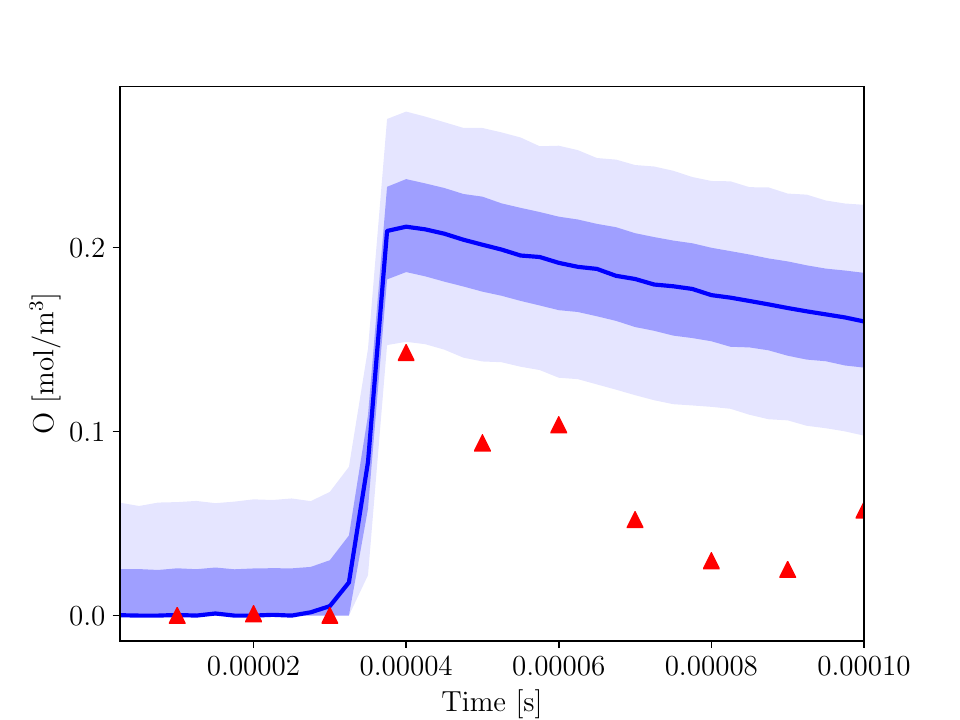}
  \end{subfigure}
  \begin{subfigure}{.49\textwidth}
    \centering
    \includegraphics[width=\textwidth]{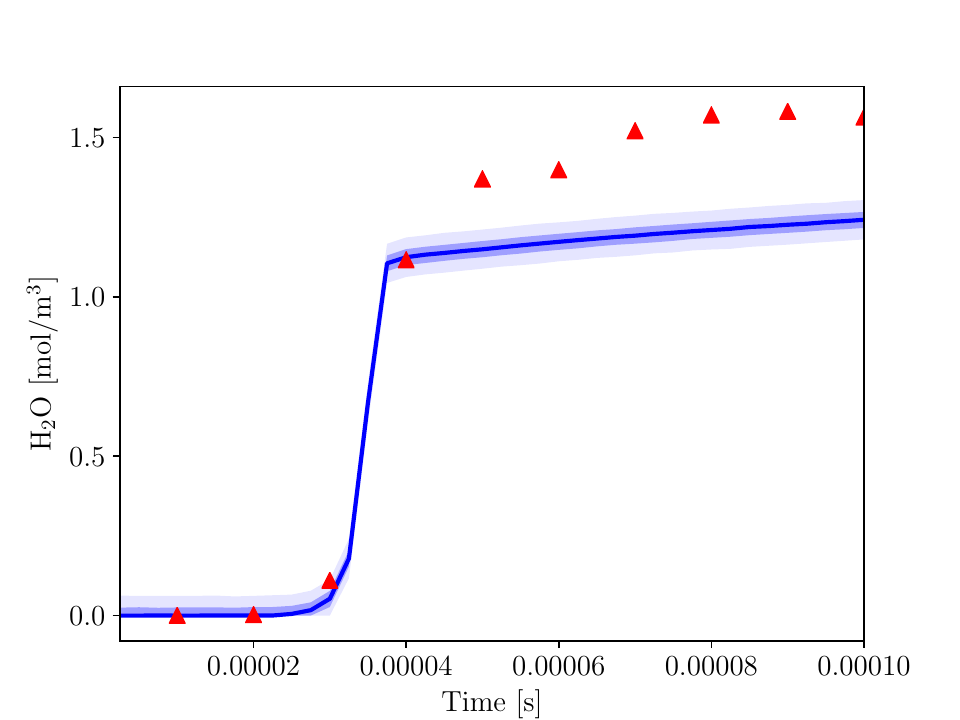}
  \end{subfigure}
  \begin{subfigure}{.49\textwidth}
    \centering
    \includegraphics[width=\textwidth]{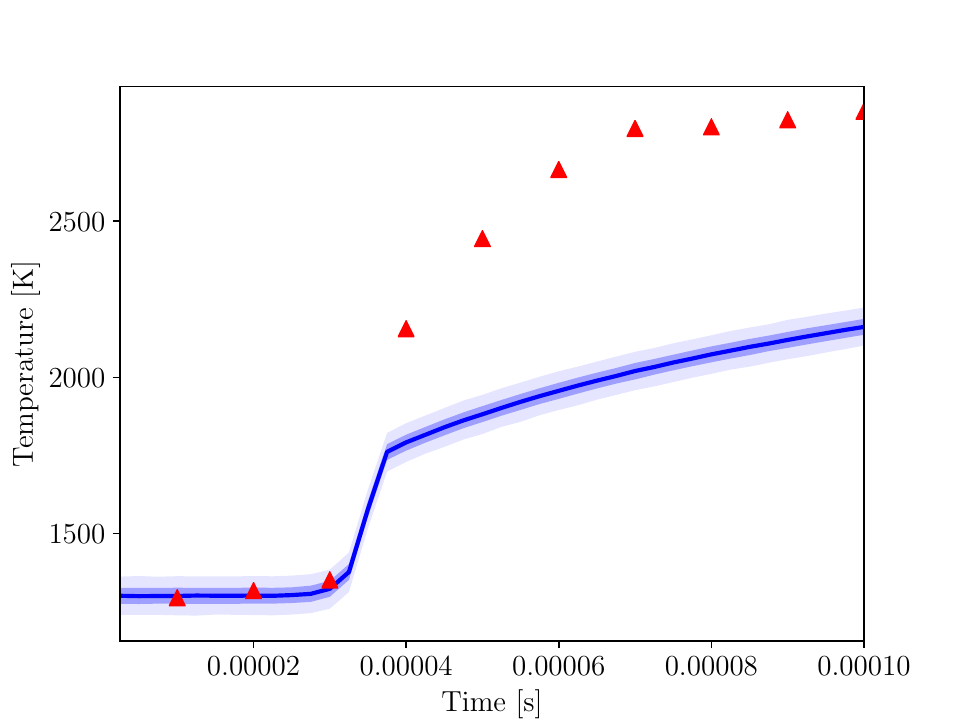}
      \caption{\label{h2-red-temp}}
  \end{subfigure}
\caption{Concentrations and temperature time-series, $\phi = 1.0$, $T_0=\SI{1300}{\kelvin}$.
Observations (red triangles), reduced model $\mathcal{R}$ (blue curves), plotted with $65$ and $95\%$ confidence intervals.
\label{fig:h2-red-smooth}}
\end{figure}

For the reduced model to be declared valid, it is required that the
model output make the observations plausible.
\revise{Figure~\ref{fig:h2-red-smooth} shows the noisy observations, generated by
the detailed model $\mathcal{D}$, compared to the reduced model
$\mathcal{R}$ output. The reduced model output is plotted with confidence intervals which account for
the measurement error.} It is clear that there are substantial
discrepancies between the reduced model and the observations for both species
concentrations and temperature.
For example, at the final observation time, the model under-predicts
the observed temperature by approximately 600K, which is far larger
than the observational uncertainty.  Further, this level of error is
not restricted to a single point.  The temperature is dramatically
under-predicted for all observations after ignition, and many of the
species predictions have errors far larger than the observational
uncertainty as well.  Thus, for any reasonable tolerance level,
inspection of figure~\ref{fig:h2-red-smooth}
shows that the reduced model alone is invalid and that a model
inadequacy representation is required if it is to be used.
\subsection{Formulation of the inadequacy operator $\mathcal{S}$}\label{sec:form}
The elements in the reduced model are H and O.  Thus, the catchall
species for the inadequacy operator are H$'$ and O$'$.  Thus, the
state for the enriched model is
\begin{align*}
\bm{x}^S = [x_1, x_2, \dots, x_9]^T,
\end{align*}
with the concentrations given in the order of H$'$, O$'$, H$_2$,
O$_2$, H, O, OH, HO$_2$, H$_2$O.  Note that $m_R = 5$, $n_R = 7$,
$n_\alpha =2$, $n_S = 9$.

\subsubsection{The random matrix $S$} 
From theorem \ref{thm:zeros}, we know that the matrix has the following structure:
\begin{equation*} S = \begin{pmatrix}
s_{1,1} & 0 & s_{1,3} & 0 & s_{1,5} & 0  & s_{1,7} & s_{1,8}& s_{1,9} \\
0 & s_{2,2} & 0 & s_{2,4} & 0 & s_{2,6}  & s_{2,7} & s_{2,8}& s_{2,9} \\
s_{3,1} & 0 & s_{3,3} & 0 & s_{3,5} & 0  & s_{3,7} & s_{3,8}& s_{3,9} \\
0 & s_{4,2} & 0 & s_{4,4} & 0 & s_{4,6}  & s_{4,7} & s_{4,8}& s_{4,9} \\
s_{5,1} & 0 & s_{5,3} & 0 & s_{5,5} & 0   & s_{5,7} & s_{5,8}& s_{5,9} \\
0 & s_{6,2} & 0 & s_{6,4} & 0 & s_{6,6}   & s_{6,7} & s_{6,8}& s_{6,9} \\
0 & 0        & 0 & 0 & 0 & 0              & s_{7,7} & s_{7,8}& s_{7,9} \\
0 & 0        & 0 & 0 & 0 & 0              & s_{8,7} & s_{8,8}& s_{8,9} \\
0 & 0        & 0 & 0 & 0 & 0              & s_{9,7} & s_{9,8}& s_{9,9} \\
\end{pmatrix}.
\label{eq:ex:S2} \end{equation*}
Here, of the 81 entries of $S$, 42 are identically zero.
Next $E$ is the $n_\alpha \times n_S$ matrix:
\begin{equation*}
E = \begin{pmatrix} 
1 & 0 & 1 & 0 & 1 & 0 & 1/2 & 1/3 & 2/3\\
0 & 1 & 0 & 1 & 0 & 1 & 1/2 & 2/3 & 1/3
\end{pmatrix}.
\end{equation*}
The first row of $E$ corresponds to hydrogen atoms and the second to oxygen.
$C$ is the following $n_S \times n_R$ matrix whose columns span $\text{null}(E)$:
\begin{equation*} C = \begin{pmatrix}
-1 & 0 & -1 & 0 & -1/2 & -1/3 & -2/3\\
0 & -1 & 0 & -1 & -1/2 & -2/3 & -1/3\\
1 & 0 & 0 & 0 & 0 & 0 & 0\\
0 & 1 & 0 & 0 & 0 & 0 & 0\\
0 & 0 & 1 & 0 & 0 & 0 & 0\\
0 & 0 & 0 & 1 & 0 & 0 & 0\\
0 & 0 & 0 & 0 & 1 & 0 & 0\\
0 & 0 & 0 & 0 & 0 & 1 & 0\\
0 & 0 & 0 & 0 & 0 & 0 & 1 \end{pmatrix}.
\label{eq:ex:C}
\end{equation*}
$P$ is an $n_R \times n_S$ matrix,
where $S = CP$:
\begin{equation*} P = 
\begin{pmatrix}
p_{1,1} & 0 & p_{1,3} & 0 & p_{1,5} & 0 &  p_{1,7} & p_{1,8} & p_{1,9}\\
0 & p_{2,2} & 0 & p_{2,4} & 0 & p_{2,6} &  p_{2,7} & p_{2,8} & p_{2,9}\\
p_{3,1} & 0 & p_{3,3} & 0 & p_{3,5} & 0 &  p_{3,7} & p_{3,8} & p_{3,9}\\
0 & p_{4,2} & 0 & p_{4,4} & 0 & p_{4,6} &  p_{4,7} & p_{4,8} & p_{4,9}\\
0 & 0 & 0 & 0 & 0 & 0 &  p_{5,7} & p_{5,8} & p_{5,9}\\
0 & 0 & 0 & 0 & 0 & 0 &  p_{6,7} & p_{6,8} & p_{6,9}\\
0 & 0 & 0 & 0 & 0 & 0 &  p_{7,7} & p_{7,8} & p_{7,9}\end{pmatrix}.
\label{eq:ex:Q}
\end{equation*}
\revise{The transformation from the entries of $P$ to $\bm{\xi}$ is given in table \ref{tab:xi2}, and the
constraints are now}
\begin{equation*} \xi_i \geq 0,\quad i=1,\dots,33.  \end{equation*}
Finally, to complete the formulation of $S$,
\begin{equation*}
\xi_i \sim \log\mathcal{N}(\mu^\xi_i, \eta^\xi_i),\quad i =1,\dots,33.
\end{equation*}

\subsubsection{The catchall reactions $\mathcal{A}$}
The catchall reactions allow the catchall species to directly form any species
made up of more than one type of atom. Otherwise, that reaction is already
allowed via $S$ ($ \cee{H$'$ -> H} $ is allowed by $S$ for example). Thus, there are three
catchall reactions:
\begin{align*}
\cee{H$'$ + O$'$ &->[$\kappa_1$] OH}\\
\cee{H$'$ + 2O$'$ &->[$\kappa_2$] HO$_2$}\\
\cee{2H$'$ + O$'$ &->[$\kappa_3$] H$_2$O}.
\end{align*}
The reaction rate coefficients are denoted $\kappa$, and these are included in
the set of inadequacy parameters. Like the variables $\bm{\xi}$, each $\kappa$
will be modeled with a lognormal distribution whose hyperparameters are also
calibrated. From the reactions above, the associated rate of each is:
\begin{align*}
  r'_1 &= \kappa_1 x_1 x_2\\
  r'_2 &= \kappa_2 x_1 x_2\\
  r'_3 &= \kappa_3 x_1 x_2,
\end{align*}
and the resulting additions to the differential equations for H$'$, O$'$, OH,
HO$_2$, and H$_2$O are
\begin{align}
  \text{H$'$:}\quad &-r'_1 - r'_2 - 2r'_3 \label{eq:a1}\\
  \text{O$'$:}\quad &-r'_1 - 2r'_2 - r'_3\\
  \text{OH:}\quad &+r'_1\\
  \text{HO$_2$:}\quad &+r'_2\\
  \text{H$_2$O:}\quad &+r'_3 \label{eq:a2}.
\end{align}
The terms above (\ref{eq:a1}) - (\ref{eq:a2}) are written as $\mathcal{A}(\bm{x})$.

\subsubsection{The energy operator $\mathcal{W}$}
The third and final piece of the operator $\mathcal{S}$ is the energy operator
$\mathcal{W}$. Recall 
\begin{equation*} \frac{dT}{dt} = \mathcal{W}(\bm{x},\dot{\bm{x}},T) = - \left(
\frac{1}{\sum_i^{n_S}{c_{v_i}(T)x_i }}\right) \left(
\sum_i^{n_S}{u_i(T) \dot{x}_i} \right) \end{equation*}
and so a description of $u(T)$ and $c_v(T)$ for each catchall species is necessary. To
do so, the new parameters $\alpha_0$, $\alpha_1$, and $\alpha_2$ are introduced. That is,
\begin{align*}
  u_i(T) &= \alpha_{0_i} + \alpha_{1_i} T + \alpha_{2_i} T^2\\
  c_{v_i}(T) &= \alpha_{1_i} + 2\alpha_{2_i} T,
\end{align*}
where $i=1$ corresponds to H$'$ and $i=2$ to O$'$.

\subsection{Calibration and validation of the inadequacy operator} \label{sec:ex_step56}
For the purposes of calibration and assessment of the stochastic
inadequacy operator, we use nine initial conditions given by the
combinations of $\phi = \{ 0.9, 1.0, 1.1\}$ and initial temperature
$T_0 = \{1200, 1300, 1400\}\SI{}{\kelvin}$. The set of time points is
again $\{20, 40, 60, 80, 100\}\SI{}{\micro \second}$, and the
observational uncertainty is as described in \S\ref{sec:ex_step23}.
The prior, likelihood, and posterior distributions exactly follow from
the general form in \S\ref{sec:calS}. Specifically, 
\begin{equation*}
    \mu^{(\cdot)}_i \sim \begin{cases}
        \mathcal{N}(0, 1e2)\,\, [1/\text{s}],\quad \text{if } (\cdot) =  \xi\\
        \mathcal{N}(0, 1e2)\,\,  [\text{cm$^3$/mol/s}] ,\quad \text{if } (\cdot) =  \kappa\\
        \mathcal{N}(0, 1e12)\,\,  [\text{J/kg/K}],\quad \text{if } (\cdot) = \alpha_0\\
        \mathcal{N}(0, 1e2)\,\,  [\text{J/kg/K$^2$}],\quad \text{if } (\cdot) =  \alpha_1\\
    \mathcal{N}(0, 1e2)\,\,  [\text{J/kg/K$^3$}],\quad \text{if } (\cdot) =  \alpha_2. \end{cases}
\end{equation*}

After calibration, the
agreement between the data and the results of the stochastic operator
model is generally very good, \revise{and is substantially better
  than with the
reduced model alone. For
example,
only 14
of the 360 observations have $\gamma\le 0.01$, and just one has $\gamma$
less than $3\times 10^{-5}$, which is the threshold suggested by the
Bonferroni correction to a 0.01 tolerance. This indicates that minor
refinement of the stochastic operator representation might be needed for
complete consistency with the data (see \S\ref{sec:con} for proposed
refinements). Nonetheless, the stochastic operator model
described here represents the primary discrepancies between the
reduced model and the high-fidelity data.}

\begin{figure}
  \centering
  \begin{subfigure}{.49\textwidth}
    \centering
    \includegraphics[width=\textwidth]{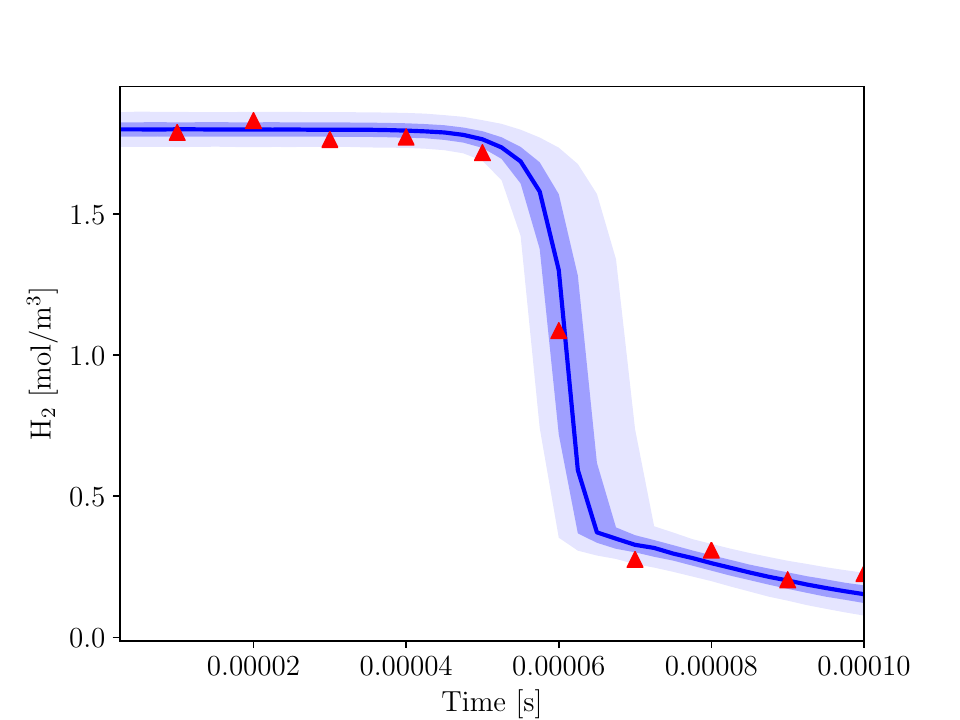}
  \end{subfigure}
  \begin{subfigure}{.49\textwidth}
    \centering
    \includegraphics[width=\textwidth]{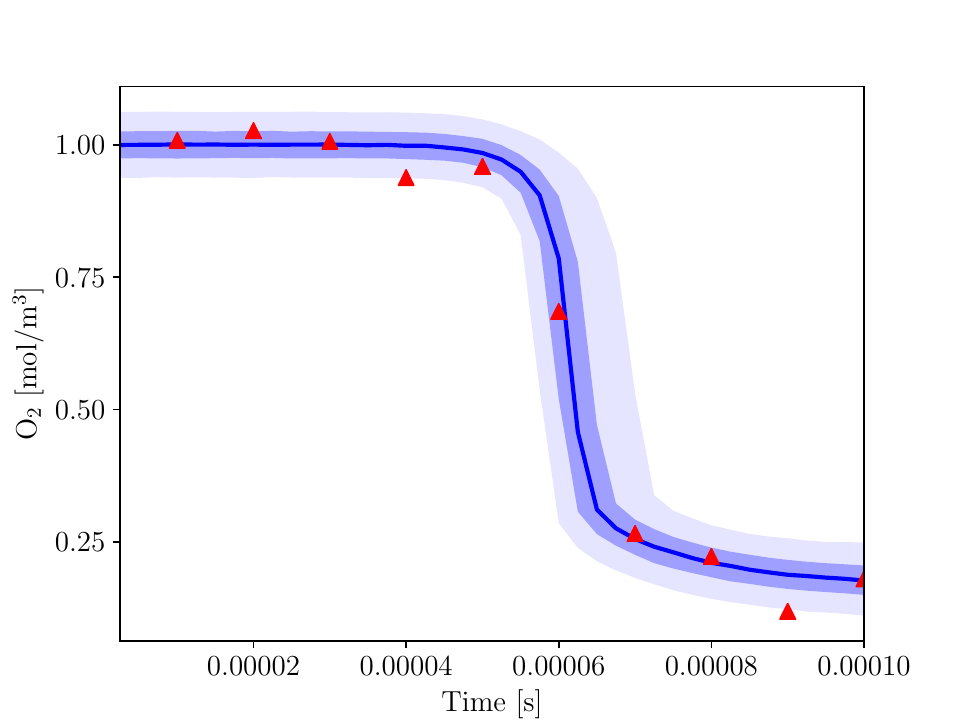}
  \end{subfigure}\\
  \begin{subfigure}{.49\textwidth}
    \centering
    \includegraphics[width=\textwidth]{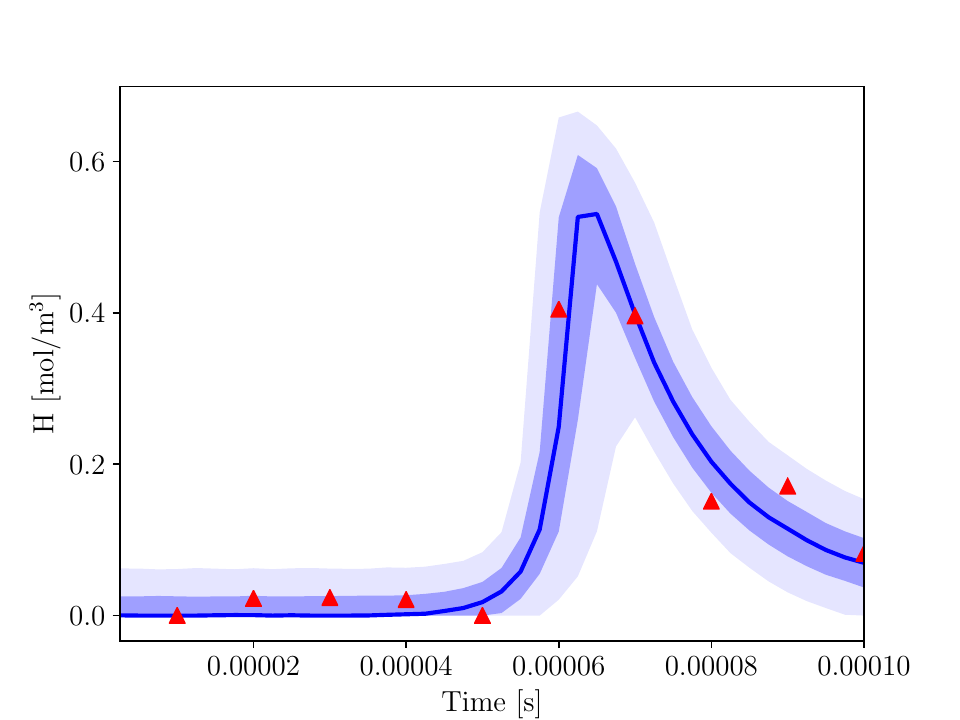}
  \end{subfigure}
  \begin{subfigure}{.49\textwidth}
    \centering
    \includegraphics[width=\textwidth]{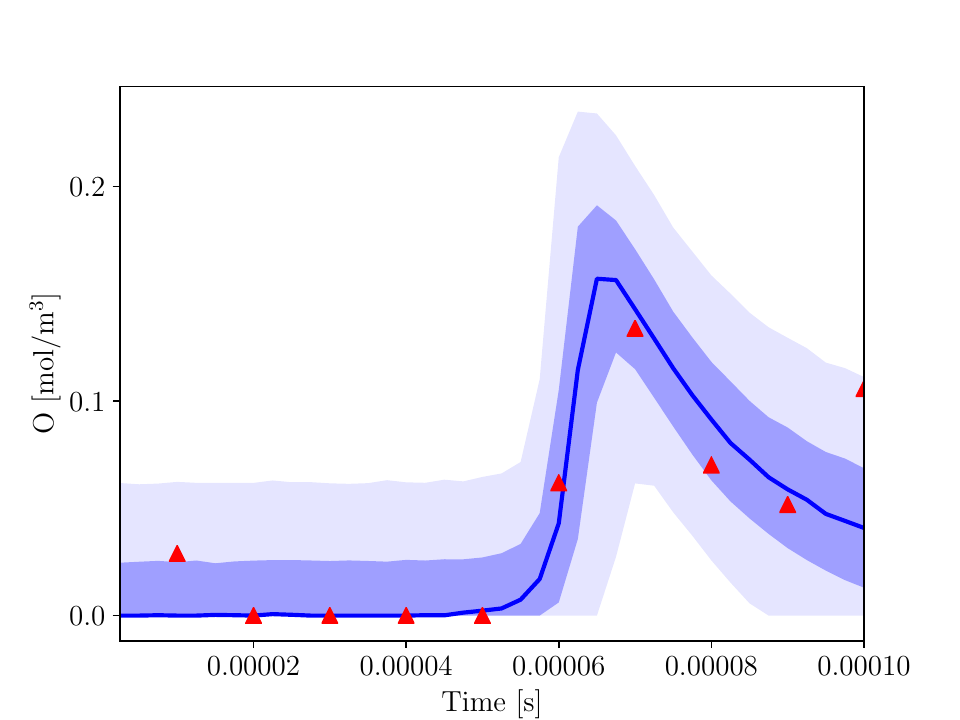}
  \end{subfigure}
  \begin{subfigure}{.49\textwidth}
    \centering
    \includegraphics[width=\textwidth]{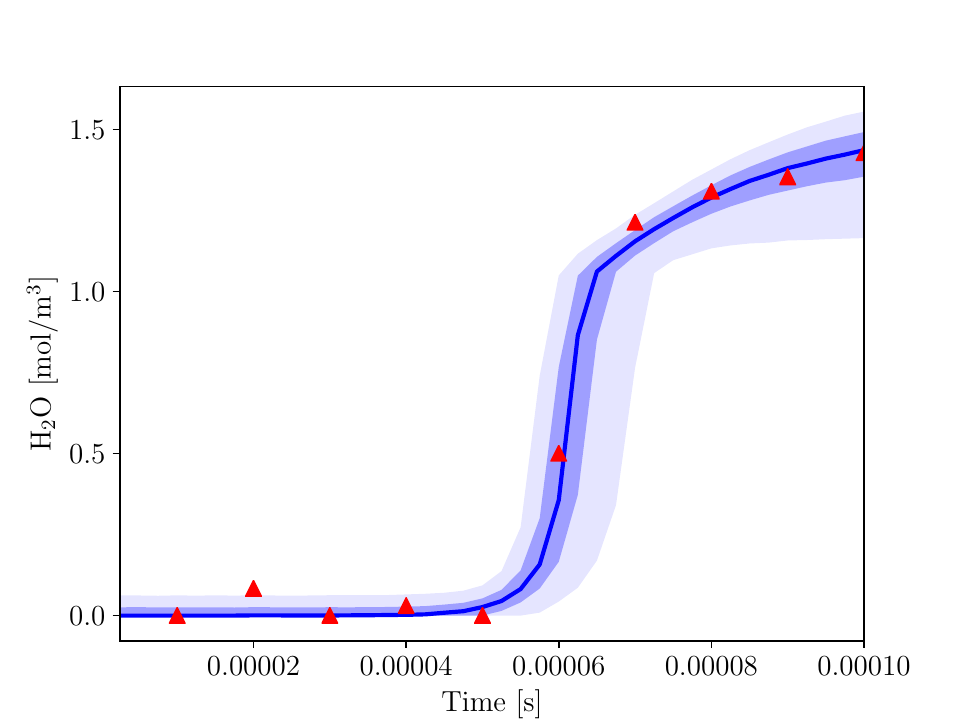}
  \end{subfigure}
  \begin{subfigure}{.49\textwidth}
    \centering
    \includegraphics[width=\textwidth]{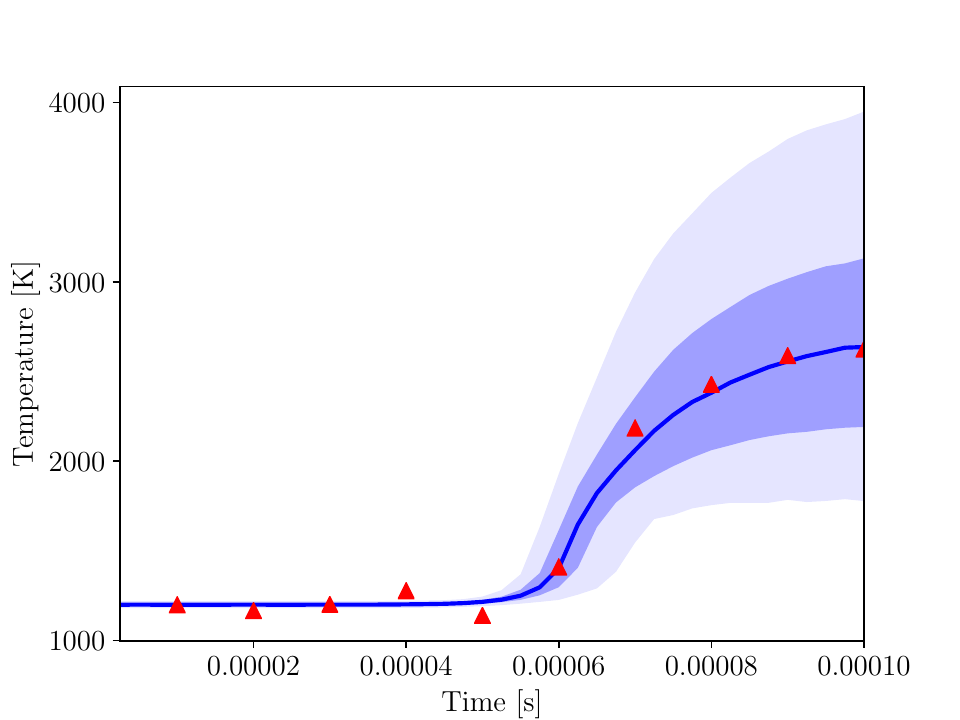}
  \end{subfigure}
\caption{Concentrations and temperature time-series,
$\phi = 0.9$, $T_0=\SI{1200}{\kelvin}$.  Observations (red triangles), reduced
  model enriched with stochastic operator inadequacy representation
 $\mathcal{O}$ (blue curves), plotted with $65$ and $95\%$ confidence intervals.
\label{fig:h2-c2-smooth0}}
\end{figure}

\begin{figure}
  \centering
  \begin{subfigure}{.49\textwidth}
    \centering
    \includegraphics[width=\textwidth]{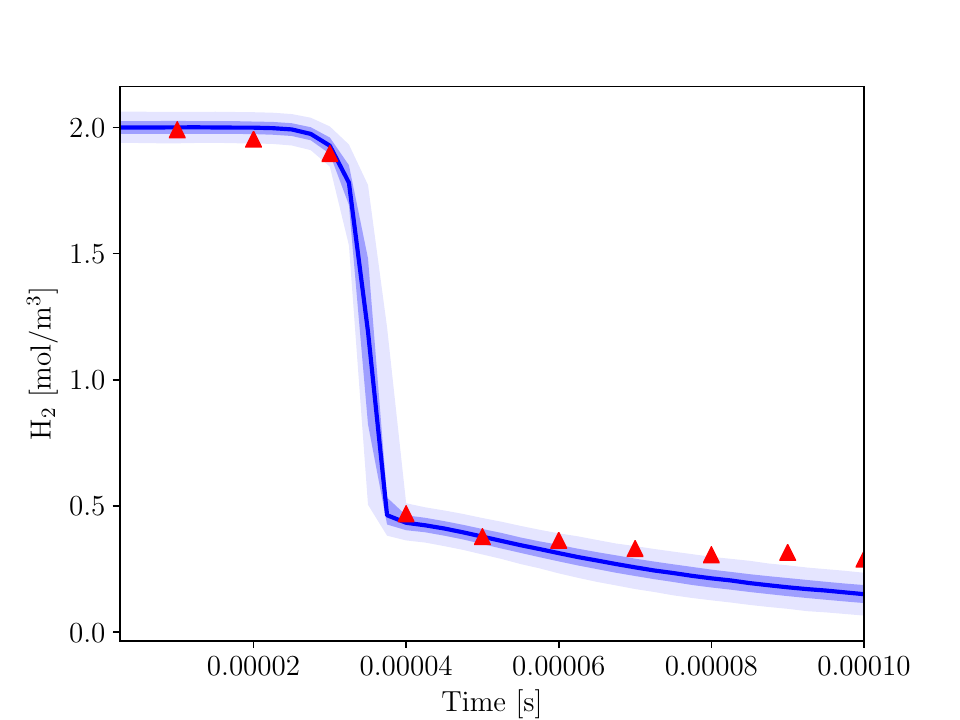}
  \end{subfigure}
  \begin{subfigure}{.49\textwidth}
    \centering
    \includegraphics[width=\textwidth]{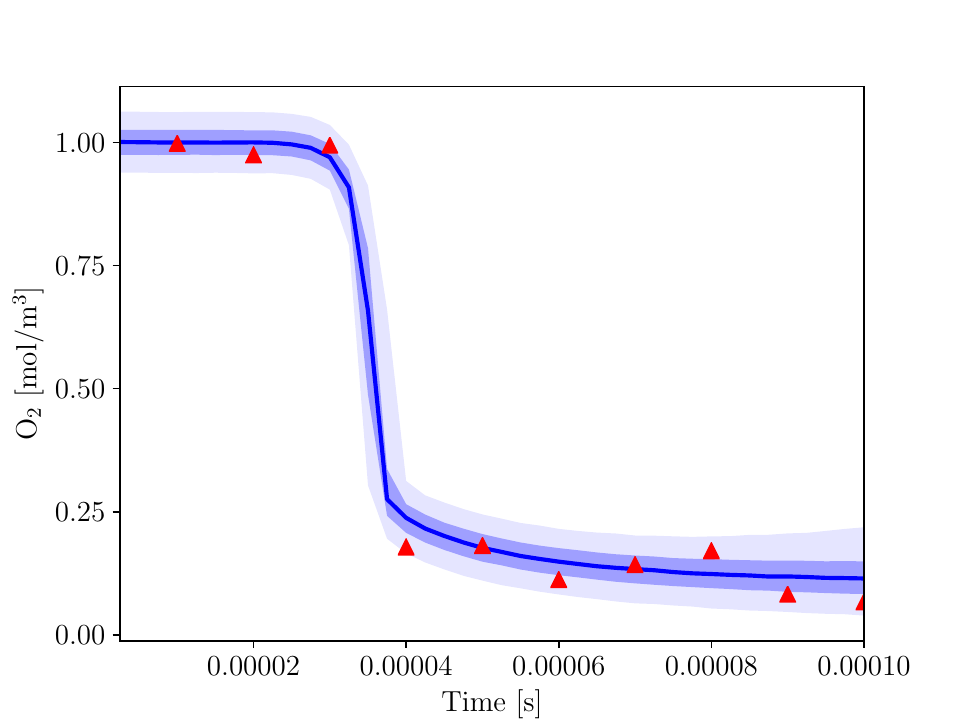}
  \end{subfigure}\\
  \begin{subfigure}{.49\textwidth}
    \centering
    \includegraphics[width=\textwidth]{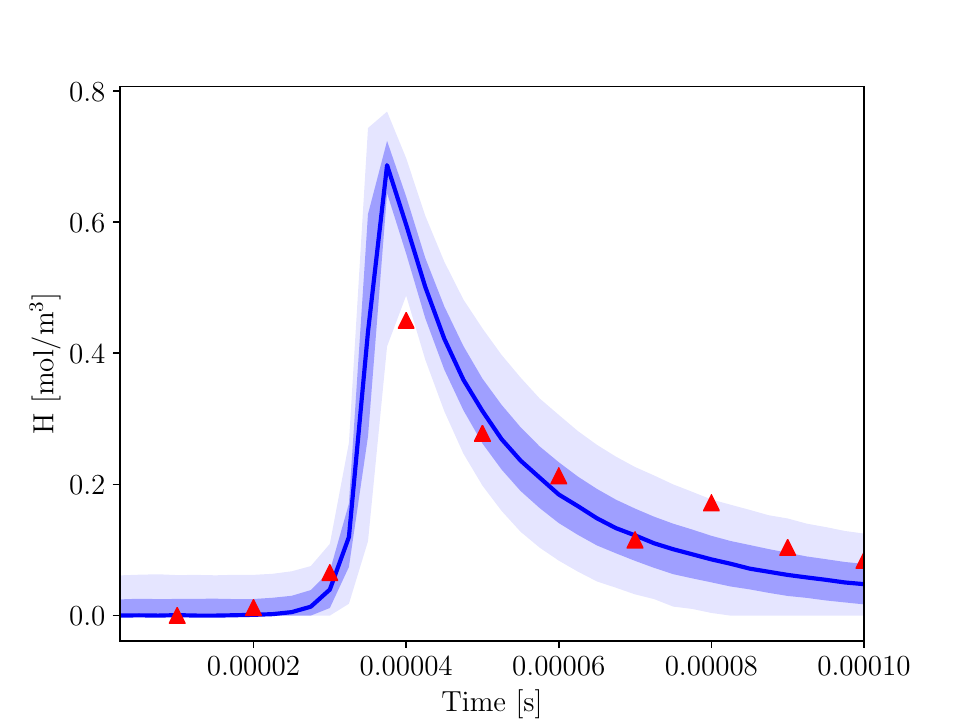}
  \end{subfigure}
  \begin{subfigure}{.49\textwidth}
    \centering
    \includegraphics[width=\textwidth]{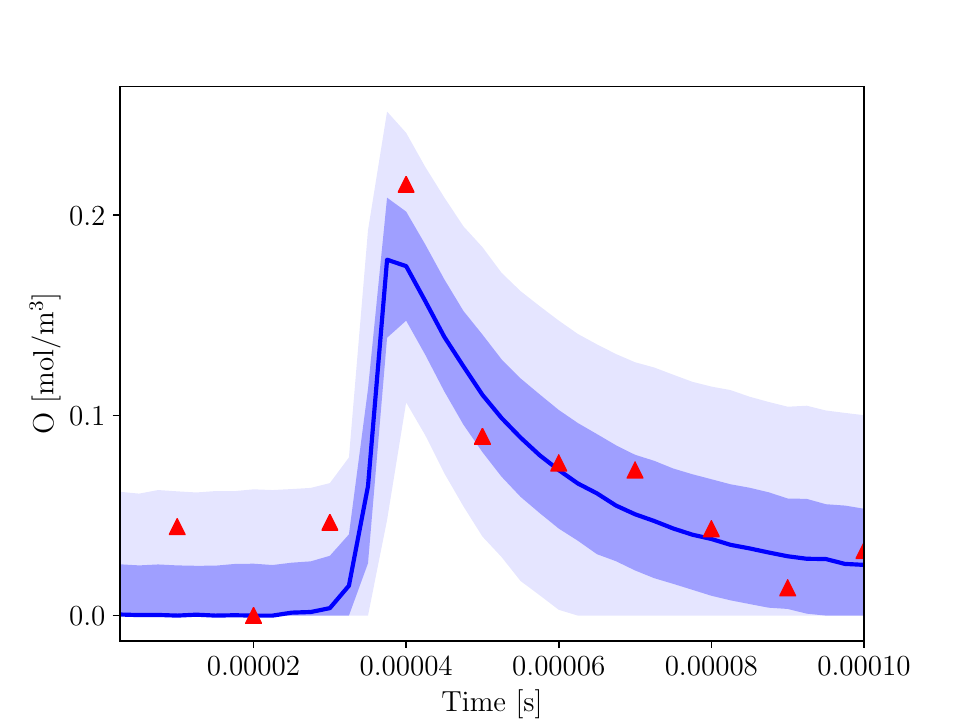}
  \end{subfigure}
  \begin{subfigure}{.49\textwidth}
    \centering
    \includegraphics[width=\textwidth]{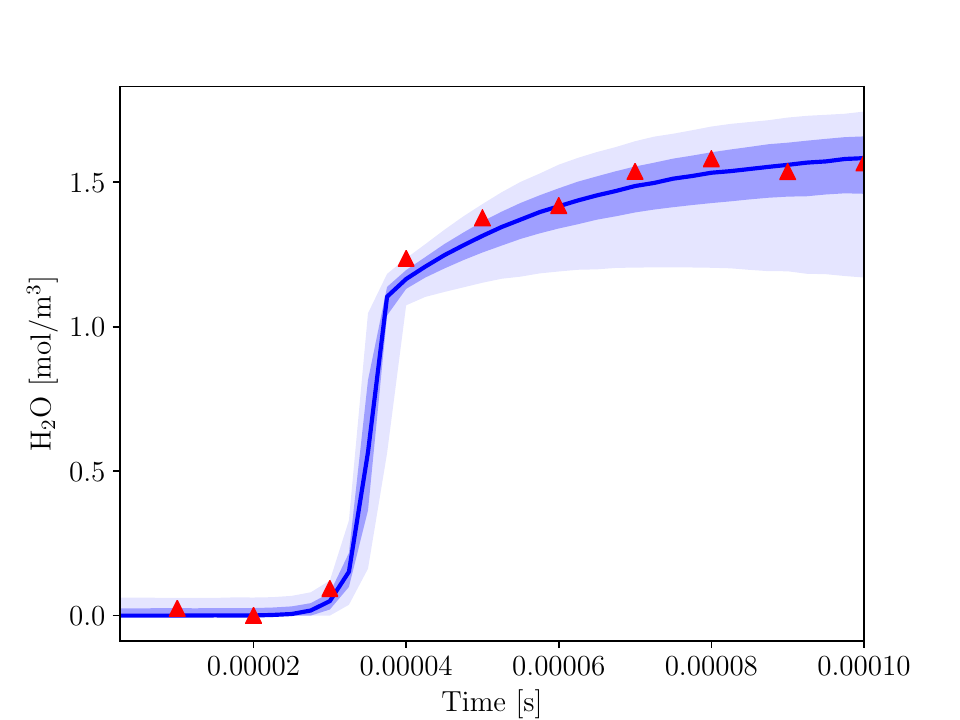}
  \end{subfigure}
  \begin{subfigure}{.49\textwidth}
    \centering
    \includegraphics[width=\textwidth]{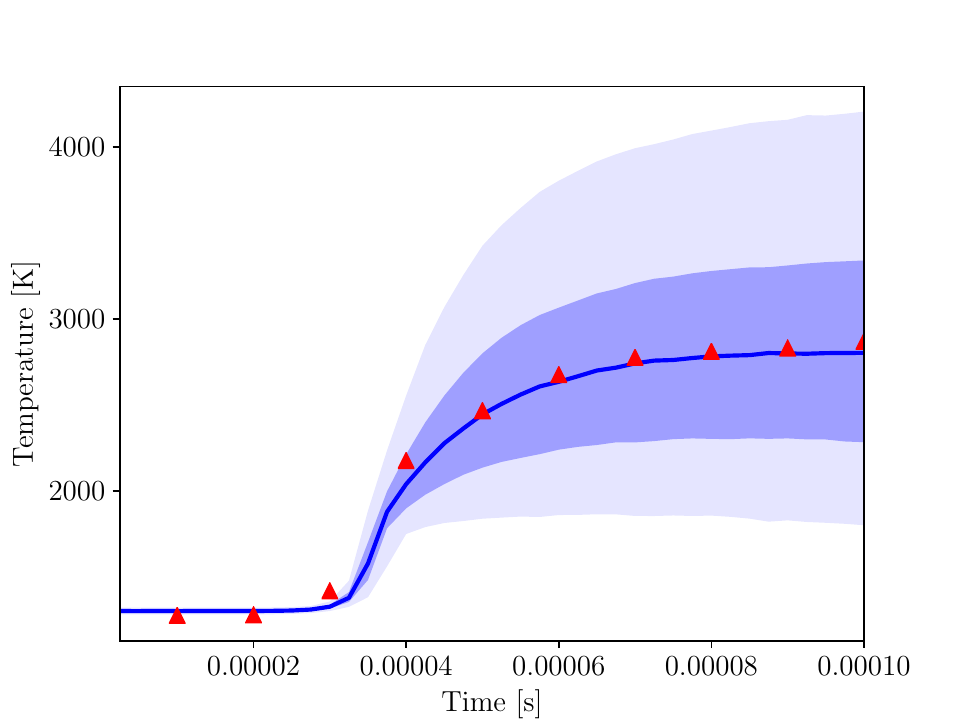}
  \end{subfigure}
\caption{Concentrations and temperature time-series, $\phi = 1.0$,
$T_0=\SI{1300}{\kelvin}$. Observations (red triangles), reduced
  model enriched with stochastic operator inadequacy representation
$\mathcal{O}$ (blue curves), plotted with $65$ and $95\%$ confidence intervals.
\label{fig:h2-c2-smooth4}}
\end{figure}

\begin{figure}
  \centering
  \begin{subfigure}{.49\textwidth}
    \centering
    \includegraphics[width=\textwidth]{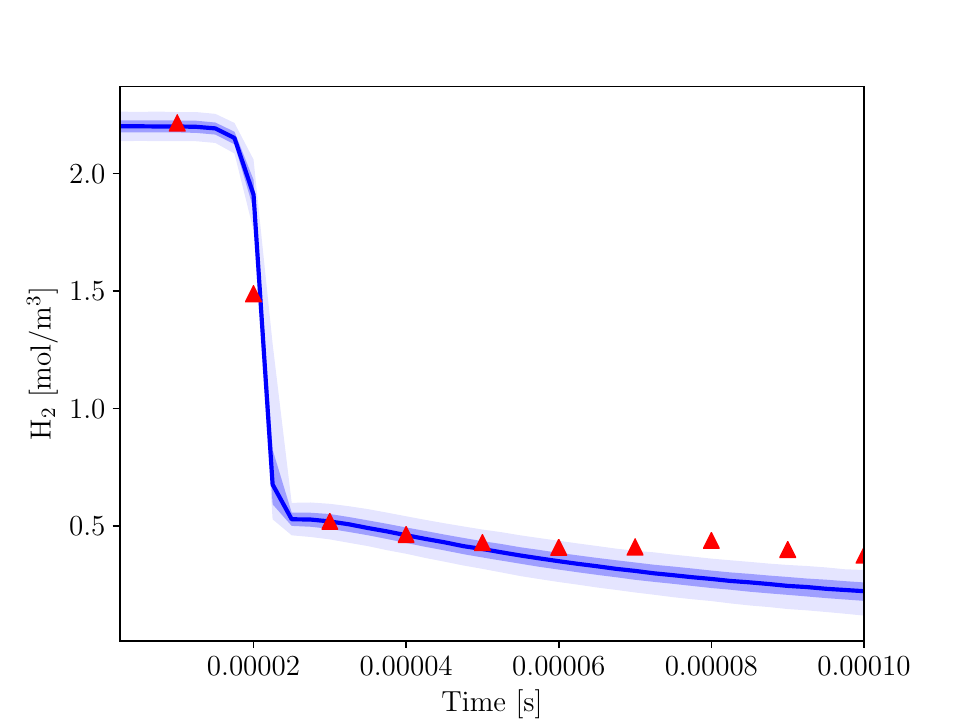}
  \end{subfigure}
  \begin{subfigure}{.49\textwidth}
    \centering
    \includegraphics[width=\textwidth]{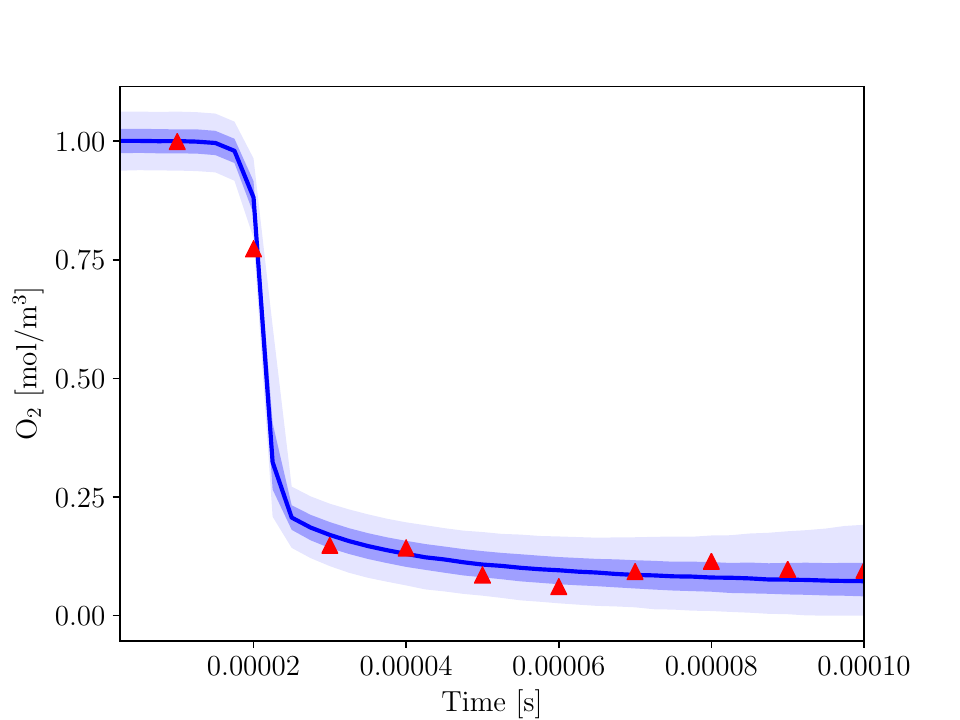}
  \end{subfigure}\\
  \begin{subfigure}{.49\textwidth}
    \centering
    \includegraphics[width=\textwidth]{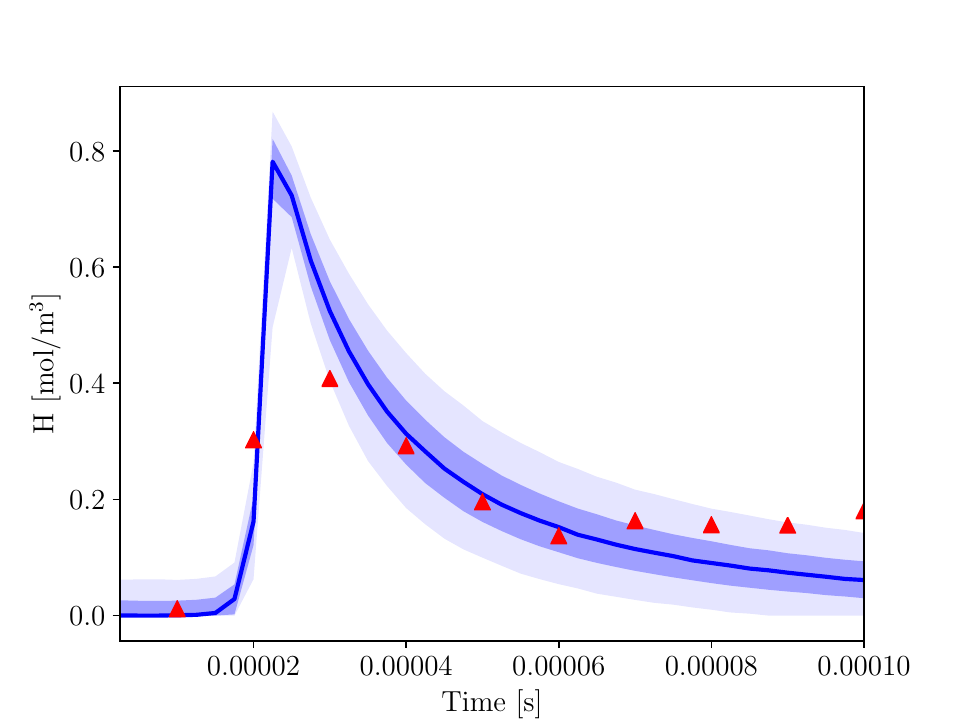}
  \end{subfigure}
  \begin{subfigure}{.49\textwidth}
    \centering
    \includegraphics[width=\textwidth]{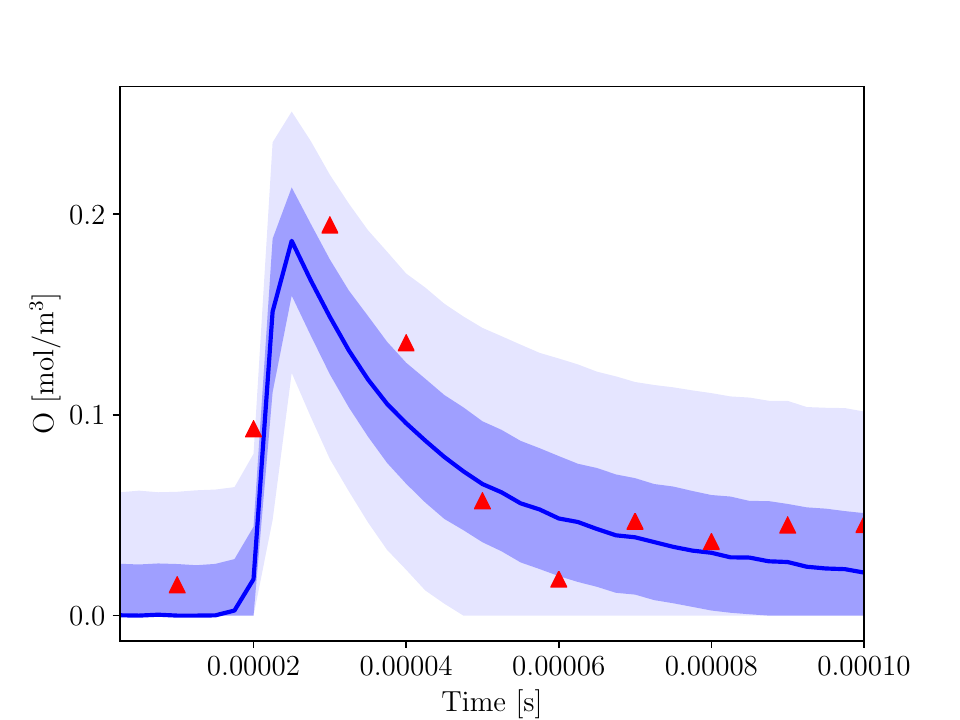}
  \end{subfigure}
  \begin{subfigure}{.49\textwidth}
    \centering
    \includegraphics[width=\textwidth]{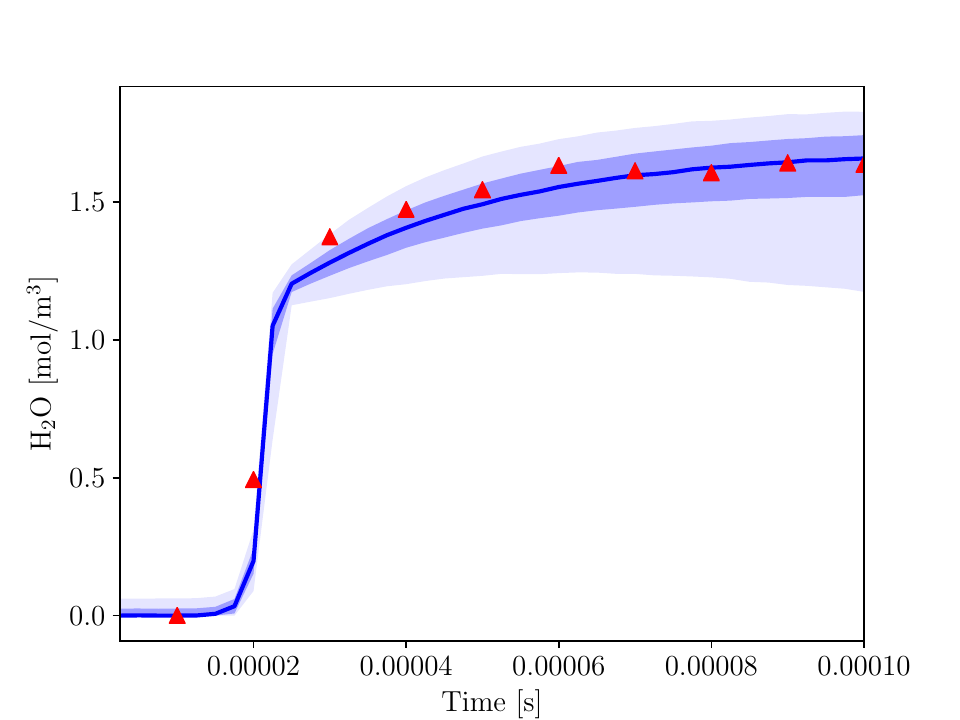}
  \end{subfigure}
  \begin{subfigure}{.49\textwidth}
    \centering
    \includegraphics[width=\textwidth]{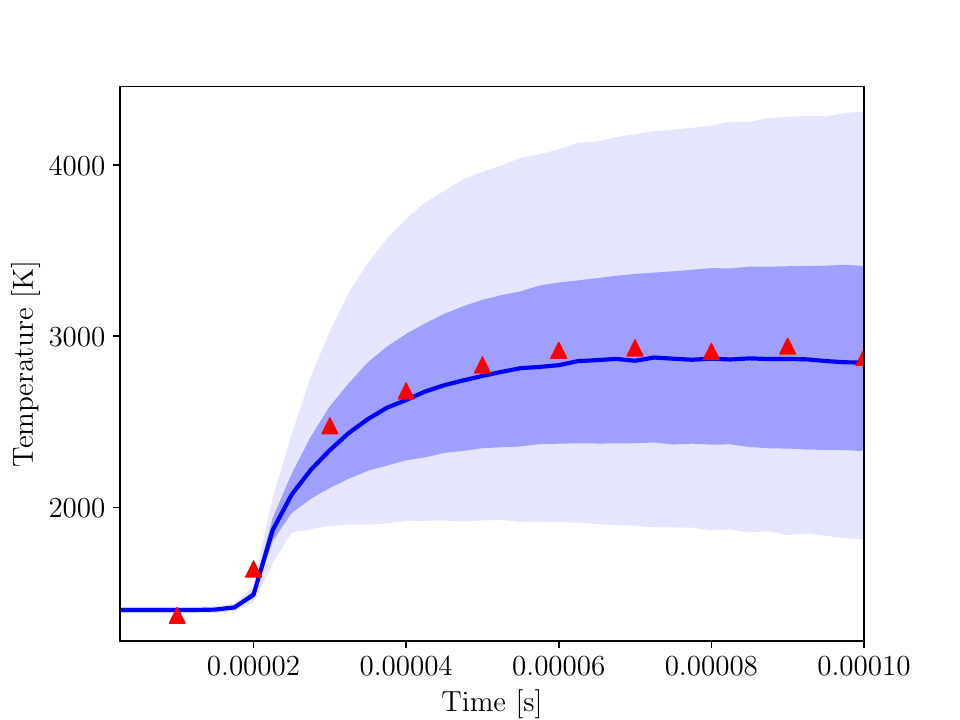}
  \end{subfigure}
\caption{Concentrations and temperature time-series, $\phi = 1.1$,
$T_0=\SI{1400}{\kelvin}$. Observations (red triangles), reduced
  model enriched with stochastic operator inadequacy representation
$\mathcal{O}$ (blue curves), plotted with $65$ and $95\%$ confidence intervals.
\label{fig:h2-c2-smooth8}}
\end{figure}

\revise{Figures~\ref{fig:h2-c2-smooth0}-\ref{fig:h2-c2-smooth8} show a representative sample of
results from the operator model
for concentrations and temperature from the nine different initial
conditions.
Note particularly the difference between figure~\ref{fig:h2-red-smooth} from the
reduced model only, and figure~\ref{fig:h2-c2-smooth4} from the stochastic operator model.
In these figures,} ten observations of the detailed model are shown in each time
series, instead of just the five that were used for calibration. This is simply to better
demonstrate the behavior of the detailed model compared to the reduced and stochastic operator
models. The observations used in the calibration correspond to the tick marks on the $x$-axis.

Consistent with the $\gamma$ results, the model prediction interval generally includes the
observations used for calibration.  The points where the observation is outside the plotted $65$ and
$95\%$ confidence intervals (and hence $\gamma$ is small) occur disproportionately at the first and
last point in time in the data set. This suggests that the operator model may have difficulty
accounting for both ignition and equilibrium behavior. Further, the mixture is initially relatively cool
compared to the observations taken later in time. Thus, the difficulty of the model capturing
both the small and large time behavior may indicate that a further enrichment of the stochastic
operator to include a temperature dependence is necessary. However, such an extension is not
mandated by the data and is beyond the
scope of the current paper.

%

\subsection{Prediction}  \label{sec:ex_step7}
We conclude this section by predicting the concentrations and
temperature for an extrapolative scenario---i.e., an initial condition
outside the range of those used to generate calibration data.  The
stochastic operator approach developed in this work has been
specifically designed with this use in mind, and it is formulated such
that there is the potential, with appropriate validation tests, that
such an extrapolative prediction could be supported by available
information.  While there are still possible improvements to the
formulation, including the temperature dependence discussed in \S
\ref{sec:ex_step56} and other possible extensions mentioned in \S
\ref{sec:con}, the fact that a validated extrapolative prediction is a
possibility distinguishes this formulations from most other inadequacy
representations. Indeed, to the authors' knowledge, this is the only
existing model inadequacy formulation for chemistry that has the
potential to make a validated prediction outside of the range of the
calibration data.

The prediction scenario shown here is a higher equivalence ratio and
lower initial temperature than included in the calibration data set.
Specifically, the initial condition is given by $\phi = 1.15$ and $T_0
= \SI{1150}{\kelvin}$.
Although the corresponding output from the detailed model was not used to calibrate the model, the
detailed model output is shown analogously to the previous results.  Since the detailed model data
are available, this extrapolation case can be viewed as an additional validation test of the
calibrated stochastic operator model.

\begin{figure}
  \centering
  \begin{subfigure}{.49\textwidth}
    \centering
    \includegraphics[width=\textwidth]{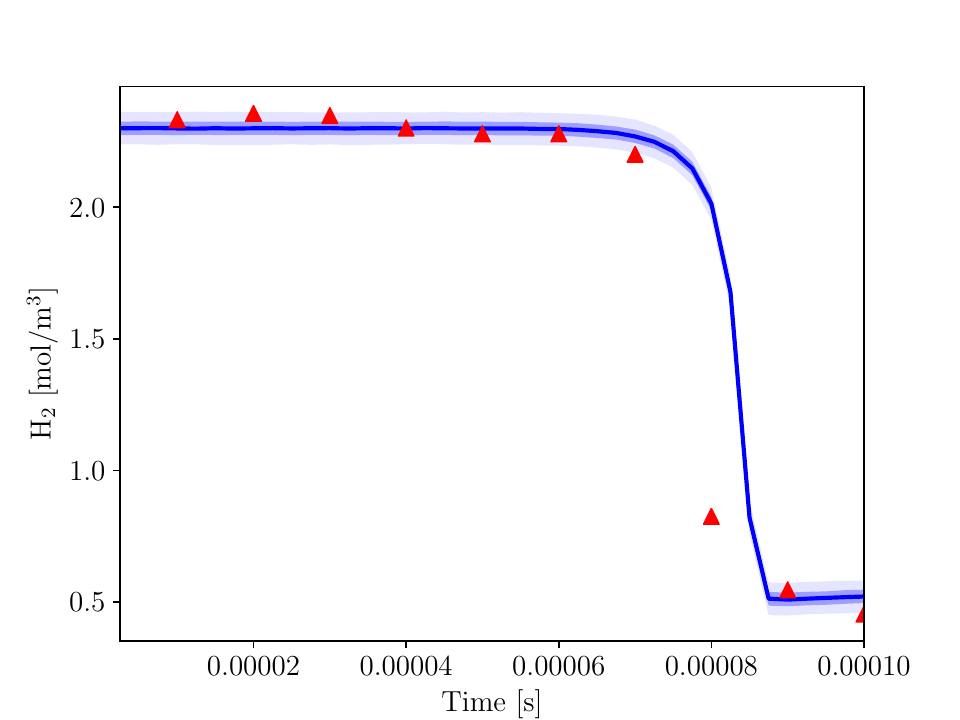}
  \end{subfigure}
  \begin{subfigure}{.49\textwidth}
    \centering
    \includegraphics[width=\textwidth]{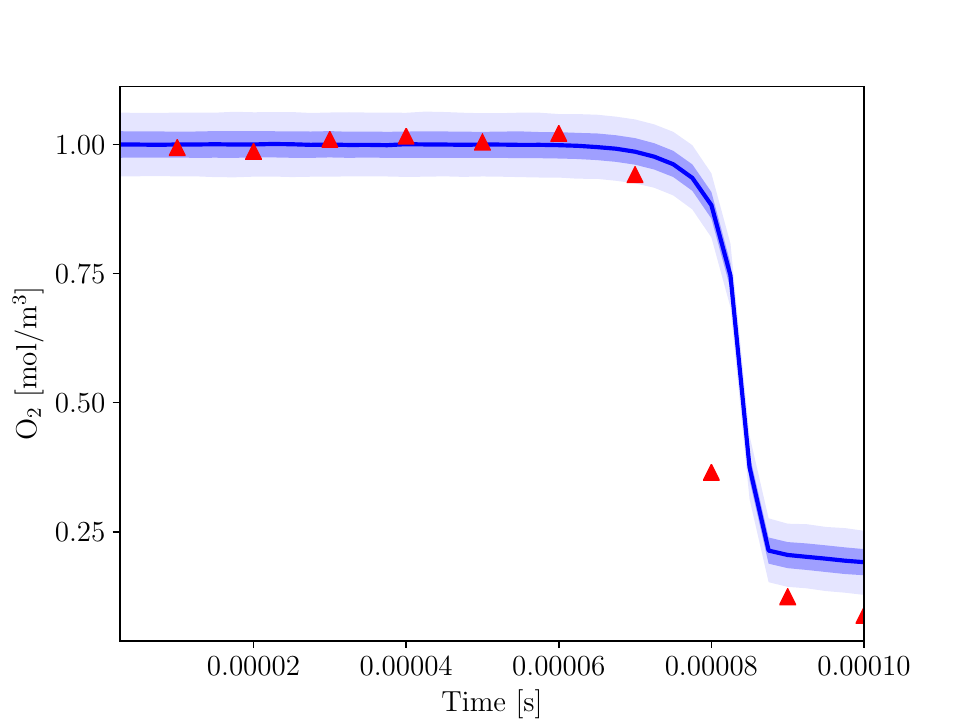}
  \end{subfigure}\\
  \begin{subfigure}{.49\textwidth}
    \centering
    \includegraphics[width=\textwidth]{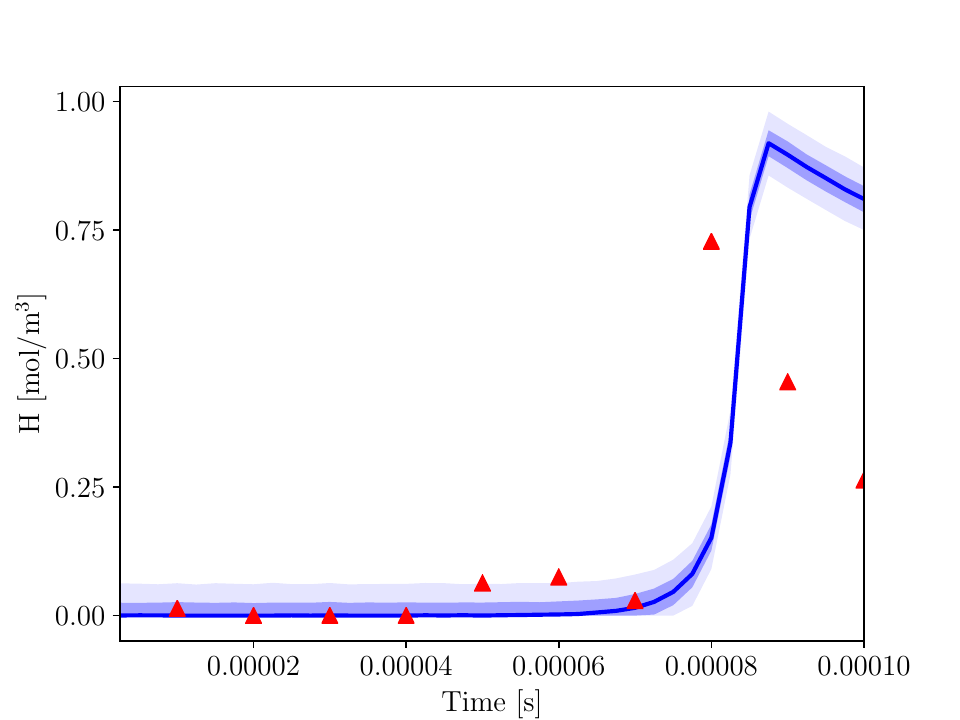}
  \end{subfigure}
  \begin{subfigure}{.49\textwidth}
    \centering
    \includegraphics[width=\textwidth]{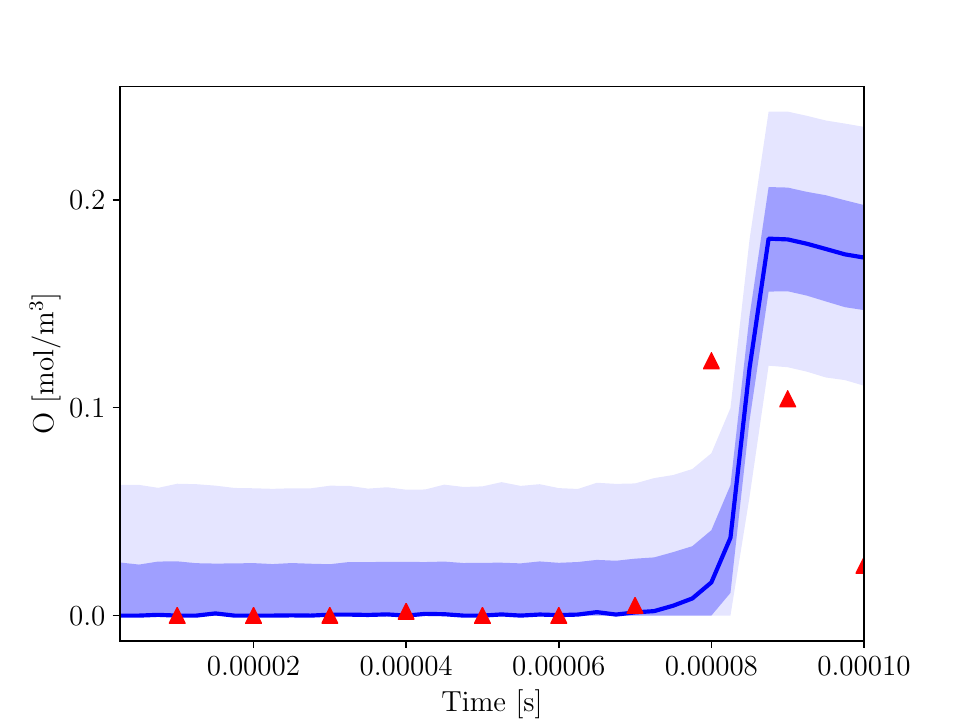}
  \end{subfigure}
  \begin{subfigure}{.49\textwidth}
    \centering
    \includegraphics[width=\textwidth]{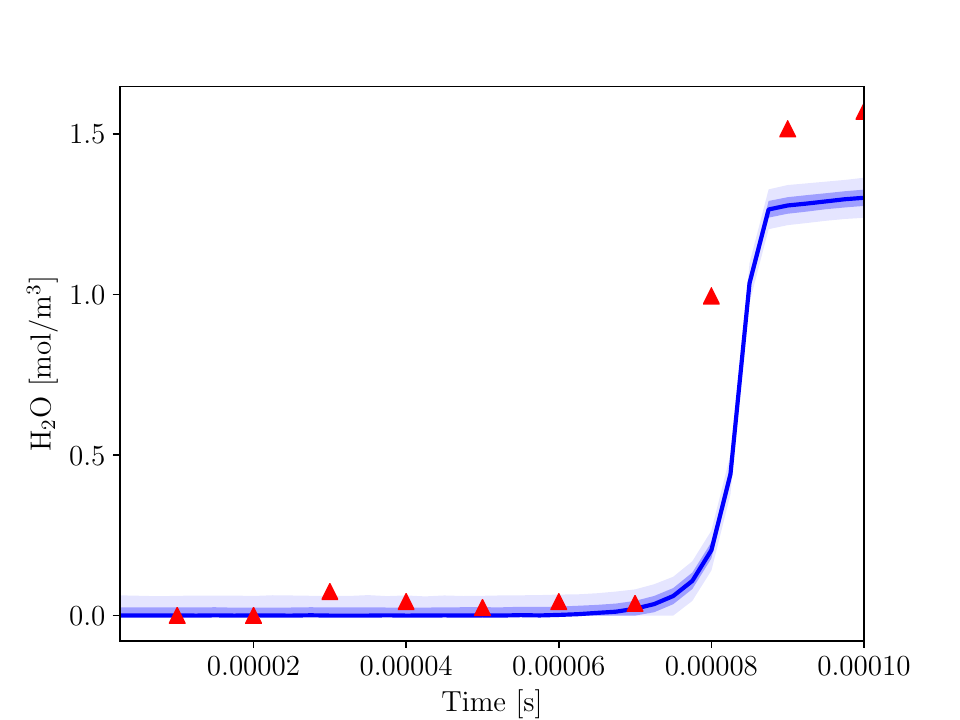}
  \end{subfigure}
  \begin{subfigure}{.49\textwidth}
    \centering
    \includegraphics[width=\textwidth]{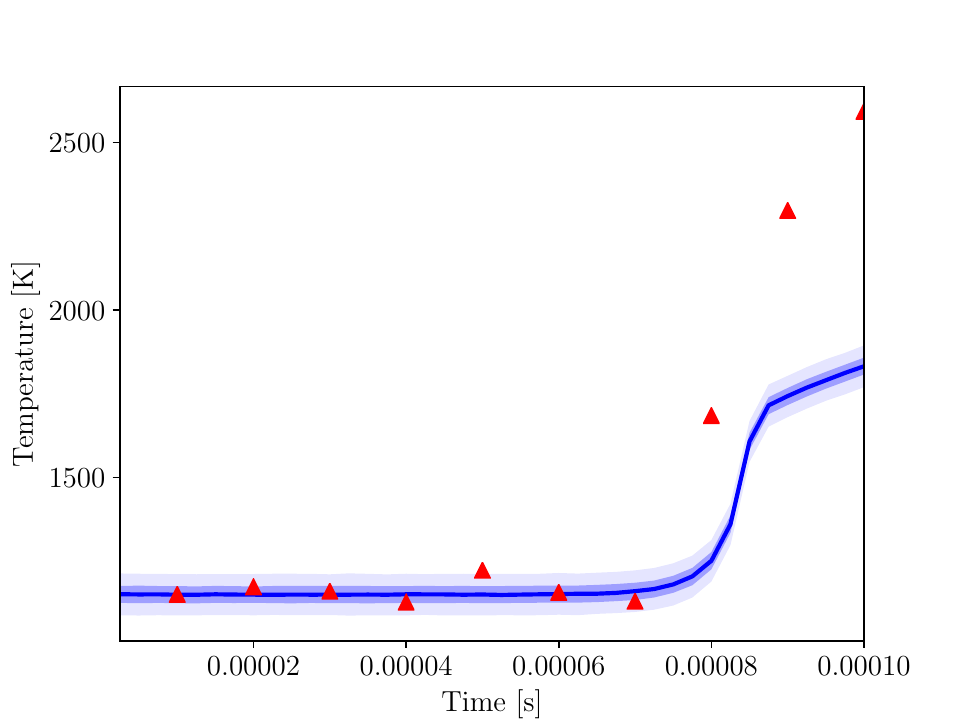}
  \end{subfigure}
\caption{Predicted concentrations and temperature time-series, $\phi = 1.15$,
$T_0=\SI{1150}{\kelvin}$. Observations (red triangles), reduced
    model (blue curves), plotted with $65$ and $95\%$ confidence intervals.
\label{fig:h2-red-smooth-pred}}
\end{figure}

For completeness of the exercise, the reduced model output for the same prediction scenario is shown
first in figure~\ref{fig:h2-red-smooth-pred}.  As expected based on
the calibration results, the reduced model alone is entirely incapable of capturing the behavior of
the system, and many of the points from the detailed model fall outside the $65$ and $95\%$
confidence intervals predicted by the reduced model.

\begin{figure}
  \centering
  \begin{subfigure}{.49\textwidth}
    \centering
    \includegraphics[width=\textwidth]{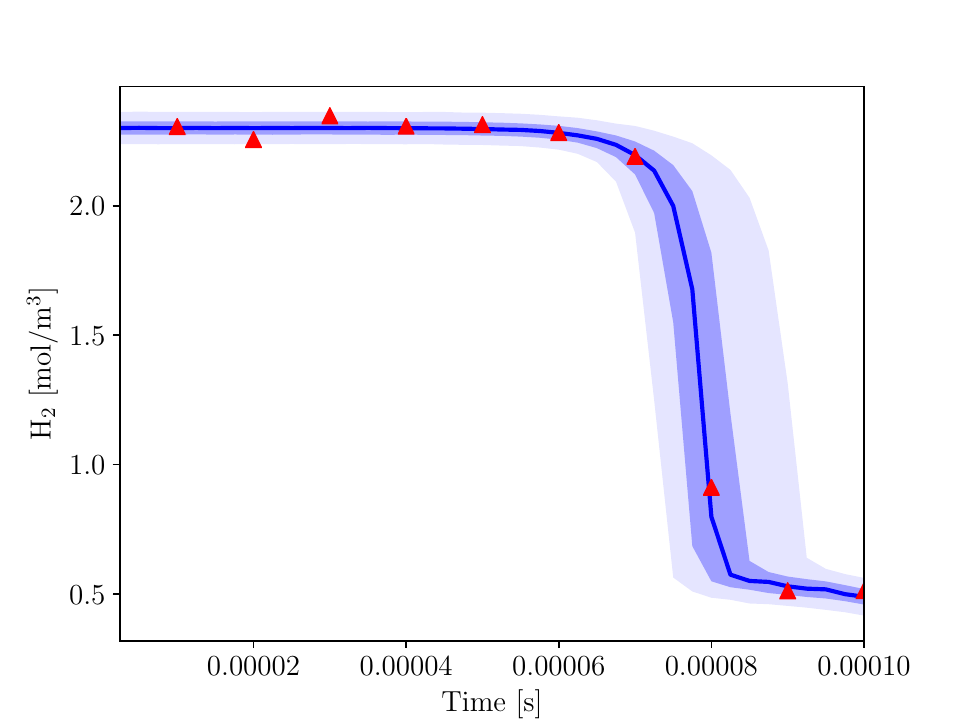}
  \end{subfigure}
  \begin{subfigure}{.49\textwidth}
    \centering
    \includegraphics[width=\textwidth]{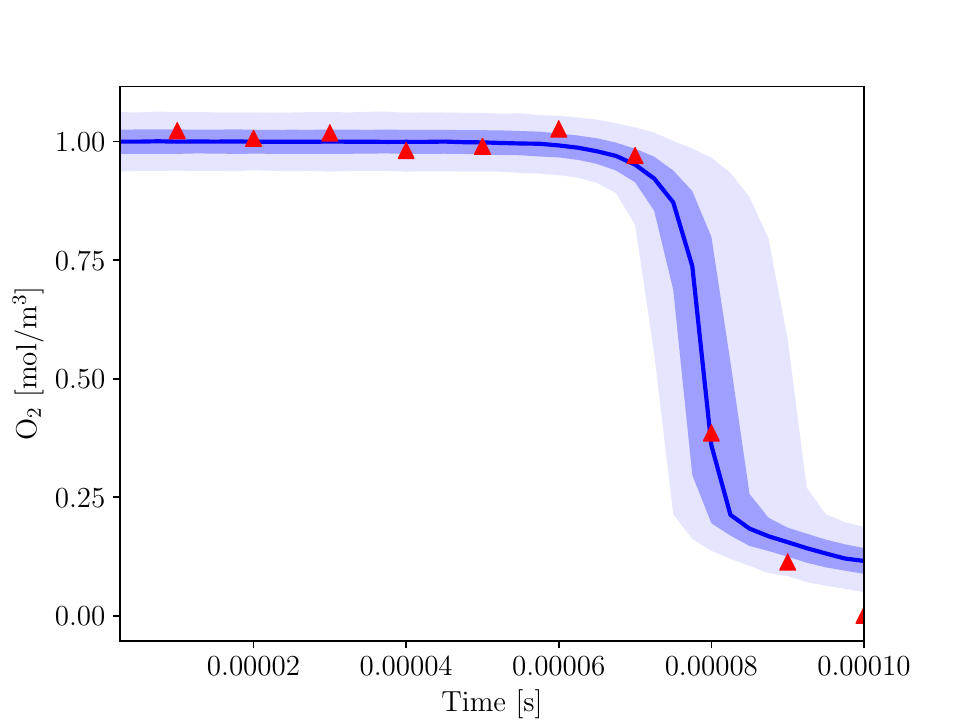}
  \end{subfigure}\\
  \begin{subfigure}{.49\textwidth}
    \centering
    \includegraphics[width=\textwidth]{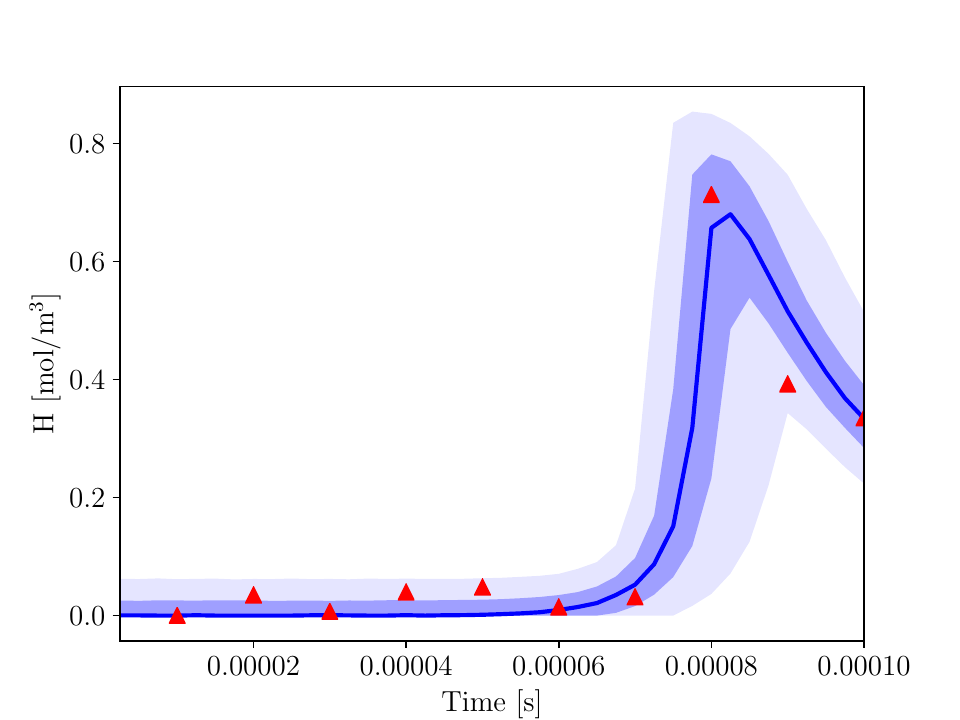}
  \end{subfigure}
  \begin{subfigure}{.49\textwidth}
    \centering
    \includegraphics[width=\textwidth]{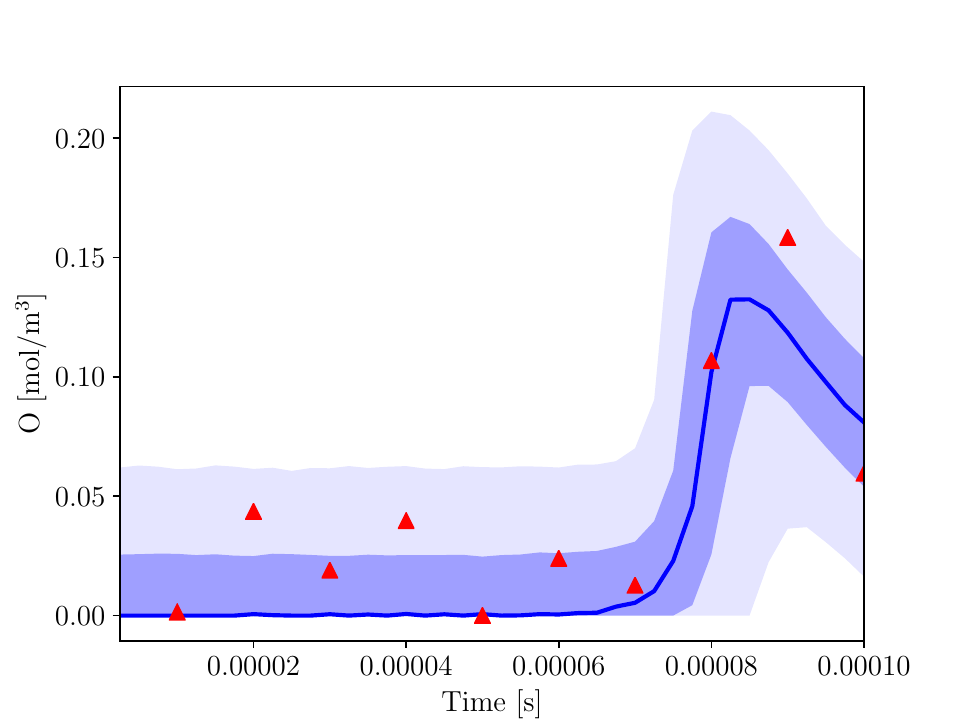}
  \end{subfigure}
  \begin{subfigure}{.49\textwidth}
    \centering
    \includegraphics[width=\textwidth]{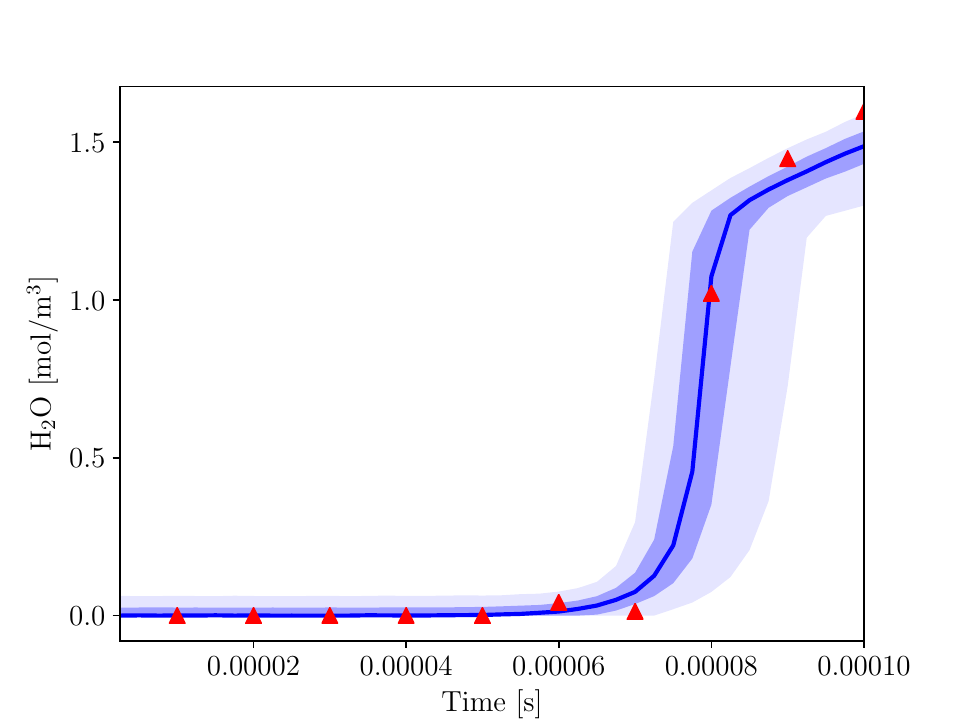}
  \end{subfigure}
  \begin{subfigure}{.49\textwidth}
    \centering
    \includegraphics[width=\textwidth]{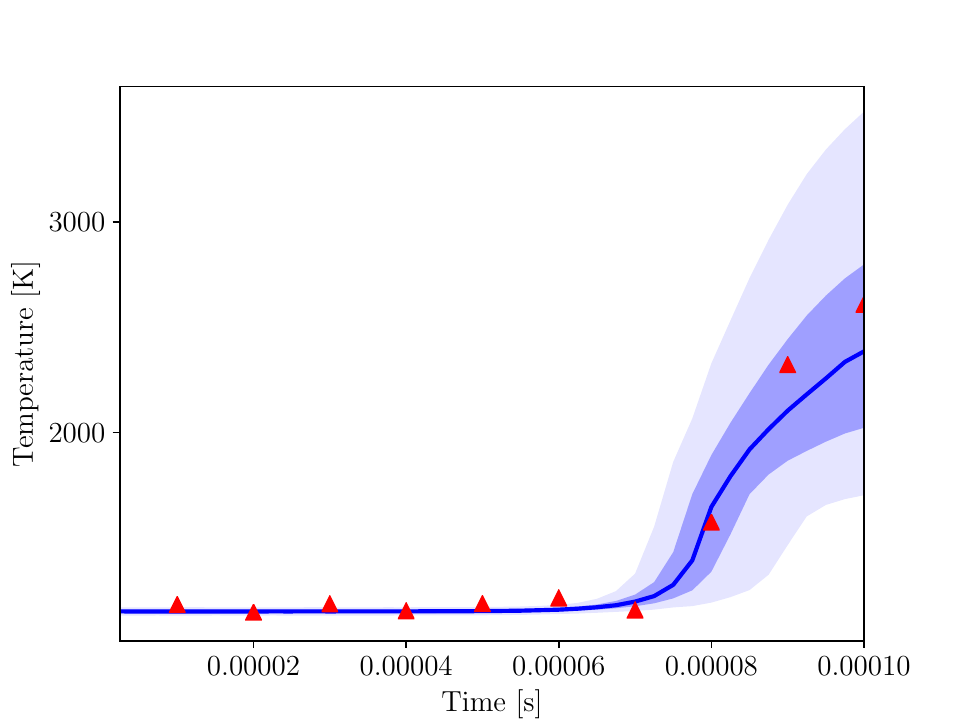}
  \end{subfigure}
\caption{Predicted concentrations and temperature time-series, $\phi = 1.15$,
$T_0=\SI{1150}{\kelvin}$. Observations (red triangles), reduced
  model enriched with stochastic operator inadequacy representation
$\mathcal{O}$ (blue curves), plotted with $65$ and $95\%$ confidence intervals.
\label{fig:h2-c2-smooth4-pred}}
\end{figure}

Figure~\ref{fig:h2-c2-smooth4-pred} shows results from the stochastic operator model.  Unlike the reduced model
alone, for the enriched model with the stochastic operator, all observations are consistent with the
prediction, \revise{with $\gamma$ values all greater than 0.01}.
%



\section{Conclusion}\label{sec:con}
This study addresses the critical problem of model inadequacy that
affects nearly all mathematical models of physical systems.  In
particular, we develop a novel approach to representing model
inadequacy in chemical kinetics models.  The approach relies on a
stochastic operator that combines the flexibility and generality of a
probabilistic model with the available deterministic physical
information. In the context of predictive models, these two properties
are essential, ensuring that the model is flexible enough to adapt to
fit available observations without generating non-physical behavior.

The stochastic operator $\mathcal{S}$ contains three main components: 1) the random
matrix $S$, 2) the nonlinear catchall reactions $\mathcal{A}$, and 3) the
energy operator $\mathcal{W}$.  The random matrix $S$ contains most of the
information in $\mathcal{S}$ and has some interesting properties. Typically, the matrix
has many identically zero entries. It always has a negative diagonal, is
diagonally dominant, and has non-positive eigenvalues. The reactions in
$\mathcal{A}$ allow any species in the reduced model to be the chemical product
of the corresponding catchall species (if this is not already possible through
$S$).  Both $S$ and $\mathcal{A}$ guarantee conservation of atoms and
non-negativity of concentrations.  Finally, the energy operator $\mathcal{W}$ modifies
the time derivative of temperature by endowing the catchall species with
internal energy.

The inadequacy operator is tested on an example H$_2$/O$_2$ mechanism.
Starting with a reduced mechanism that was shown to be invalid for
prediction, the stochastic operator model is able to account for the
model inadequacy.  The observations used for
calibration are plausible outcomes of the operator model---i.e., the
concentrations and temperature values from the operator model
$\mathcal{O}$ are consistent with the data supplied by the detailed
model $\mathcal{D}$. Moreover, the prediction case showed that the
operator model output was consistent with all of the observations
from the detailed model for a scenario outside the range of initial
conditions used for the operator calibration.

There are many avenues for extending the work reported here. Ongoing
research includes the application of the operator to a more complex
chemical setting, namely, a methane-air mechanism.  We are also
investigating a more complex physical setting with a more complex prediction
problem: the prediction of a hydrogen laminar flame using the
calibrated operator found in this work. Consistent with the discussion
in \S\ref{sec:ex_step56}, preliminary research on these topics has
suggested that the temperature-independence of the operator (save
$\mathcal{W}$) is too severe a limitation, and that in complex
problems, the more information incorporated into the stochastic
operator, the better. Thus, immediate goals are to include more
physical information in the inadequacy representation such as
realistic temperature-dependence, and to use stronger priors based on
knowledge of the chemical reactions and the physical setup. This leads
to the next major opportunity for future work: developing the
connection between the stochastic operator and the actual
chemistry. Mapping between the random matrix and the typical chemical
reactions was a first step in this direction. However, a better
understanding of what the stochastic operator means in physical terms
is needed.  This includes not just the structure, but also the
uncertainty in the calibrated parameters. A future goal is to infer
something about the missing chemistry from the calibrated operator.
This is important when developing mechanisms based on experimental
data, when no detailed mechanism is available.

Further, it is unclear that the random matrix coupled with catchall
species and reactions is the optimal way to formulate the stochastic
operator. \revise{The inadequacy formulation introduces many
  calibration parameters, which in turn increase the computational
  complexity. While the increase in dimensionality was manageable in
  this example problem, it may become impractical in more complex
  problems, especially when the reduced kinetics model is more
  complex.} A number of alternative formulations are being
pursued. For example, instead of using the random matrix $S$ and the
catchall reactions in $\mathcal{A}$, a simplified version is the
following: given $n_R$ species in the reduced model, include $n_R$
reversible dissociation reactions where the reactant is one of the
original species and the products are the corresponding catchall
species. In the reverse combination reaction, the catchall atoms react
to form any of the original species. In this fashion, the atoms of any
species could move to any other species in two steps. This
representation would lose the flexibility possible in the current
formulation; that is, the current formulation allows for the most
detailed and direct linear movement from one species to another. If
any such direct pathway proves significant, this would not be captured
by this simplified formulation.  On the other hand, it would decrease
the number of random variables and thus would be more tractable in
more complex reaction systems.  With a smaller number of additional
reactions, the corresponding reaction rates could then be enriched
with temperature-dependence.

Another variation could be a more complete set of nonlinear reactions. Instead
of only allowing the nonlinear catchall reactions, one could augment the reduced
model with all or some subset of all possible nonlinear terms. In contrast to
the first variation, this would increase the number of random variables. A
formulation like this might only be possible with more informative priors or
knowledge about the chemical system.

Finally, it would be very informative to apply this method to new
problems.  For instance, the inadequacy operator could be tested by a
more realistic combustion problem. It may be that doing so requires a
more complete thermodynamic description of the catchall species.  More
importantly, there is no reason to restrict the stochastic operator
approach developed here to models of chemically reacting gas mixtures.
Similar model structures and inadequacies appear in many other
modeling domains.  The guiding principles of this work (respecting
physical constraints, maintaining flexibility, starting with a
linearized version) and developing an analogous operator (possibly
random matrix) should be explored in many different physical settings.
Applications in many different domains could bring to light new
challenges and common strengths for the stochastic operator approach
to representing model inadequacy.

\newpage
\begin{appendices}
\setcounter{table}{0}
\numberwithin{table}{section}
\renewcommand*\thetable{\Alph{section}.\arabic{table}}

\section{Reaction mechanisms}\label{app:mech}
The 21 reactions in the detailed hydrogen-oxygen mechanism are listed in table
\ref{tab:21rxn} and the five of the reduced mechanism in \ref{tab:5rxn}. The
associated reaction rate is $k = A T^b e^{-E/R^{\circ}T}$.
\begin{table}
\begin{centering}
\begin{tabular}{l l l l}
\hline
Reaction & $A$ & $b$ & $E$ \\
\hline
{\it Hydrogen-oxygen chain} & & & \\
1. \cee{H + O2 -> OH + O} & $3.52 \times 10^{16}$ & -0.7 & 71.4 \\
2. \cee{H2 + O -> OH + H} & $5.06 \times 10^4$    &  2.7 & 26.3 \\
3. \cee{H2 + OH -> H2O + H} & $1.17 \times 10^9$    & 1.3 & 15.2 \\
4. \cee{H2O + O -> OH + OH} & $7.60 \times 10^0$    & 3.8 & 53.4 \vspace{.2cm}\\
{\it Direct recombination} & & & \\
5. \cee{H + H + M -> H2 + M} & $1.30 \times 10^{18}$ & -1.0 & 0.0 \\
6. \cee{H + OH + M -> H2O + M} & $4.00 \times 10^{22}$ & -2.0 & 0.0 \\
7. \cee{O + O + M -> O2 + M} & $6.17 \times 10^{15}$ & -0.5 & 0.0 \\
8. \cee{H + O + M -> OH + M} & $4.71 \times 10^{18}$ & -1.0 & 0.0 \\
9. \cee{O + OH + M -> HO2 + M} & $8.00 \times 10^{15}$ & 0.0 & 0.0 \vspace{.2cm}\\
{\it Hydroperoxyl reactions} & & & \\
10. \cee{H + O2 + M -> HO2 + M} & $5.75 \times 10^{19}$ & -1.4 & 0.0 \\
11. \cee{HO2 + H -> OH + OH} & $7.08 \times 10^{13}$ & 0.0 & 1.2 \\
12. \cee{HO2 + H -> H2 + O2} & $1.66 \times 10^{13}$ & 0.0 & 3.4 \\
13. \cee{HO2 + H -> H2O + O} & $3.10 \times 10^{13}$ & 0.0 & 7.2 \\
14. \cee{HO2 + O -> OH + O2} & $2.00 \times 10^{13}$ & 0.0 & 0.0 \\
15. \cee{HO2 + OH -> H2O + O2} & $2.89 \times 10^{13}$ & 0.0 & -2.1 \vspace{.2cm}\\
{\it Hydrogen peroxide reactions} & & & \\
16. \cee{OH + OH + M -> H2O2 + M} & $2.30 \times 10^{18}$ & -0.9 & -7.1 \\
17. \cee{HO2 + HO2 -> H2O2 + O2} & $3.02 \times 10^{12}$ & 0.0 & 5.8 \\
18. \cee{H2O2 + H -> HO2 + H2} & $4.79 \times 10^{13}$ & 0.0 & 33.3 \\
19. \cee{H2O2 + H -> H2O + OH} & $1.00 \times 10^{13}$ & 0.0 & 15.0 \\
20. \cee{H2O2 + OH -> H2O + HO2} & $7.08 \times 10^{12}$ & 0.0 & 6.0 \\
21. \cee{H2O2 + O -> HO2 + OH} & $9.63 \times 10^6$    & 2.0 & 2.0 \\
\hline
\end{tabular}\\
\begin{tabular}{c} Units: mol, cm, s, kJ, K.
\end{tabular} 
\caption{The detailed H$_2$/O$_2$ reaction mechanism from \cite{williams08}. \label{tab:21rxn}}
\end{centering}
\end{table}

\begin{table}
\begin{centering}
\begin{tabular}{l l l l}
\hline
Reaction & $A$ & $b$ & $E$ \\
\hline
{\it Hydrogen-oxygen chain} & & & \\
1. \cee{H + O2 -> OH + O} & $3.52 \times 10^{16}$ & -0.7 & 71.4 \\
2. \cee{H2 + O -> OH + H} & $5.06 \times 10^4$    &  2.7 & 26.3 \\
3. \cee{H2 + OH -> H2O + H} & $1.17 \times 10^9$    & 1.3 & 15.2\vspace{.2cm} \\
{\it Hydroperoxyl reactions} & & & \\
10. \cee{H + O2 + M -> HO2 + M} & $5.75 \times 10^{19}$ & -1.4 & 0.0 \\
12b. \cee{H2 + O2 -> HO2 + H} & $1.4 \times 10^{14}$ & 0.0 & 249.5 \\
\hline
\end{tabular}\\
\begin{tabular}{c} Units: mol, cm, s, kJ, K.
\end{tabular} 
\caption{The reduced H$_2$/O$_2$ reaction mechanism from \cite{williams08}.\label{tab:5rxn}}
\end{centering}
\end{table}

\section{Properties of $S$} \label{app:Sprops}

\subsection{Sparsity}
As noted in \S\ref{sec:sparse}, $S$ is often sparse.  There is a way
to determine which entries of $S$ are identically zero using the
physical restrictions about how different species concentrations
interact with each other. To determine the sparsity in this way, each
species $\mathbb{X}$ is characterized by a composite number
$\rho_{\mathbb{X}}$. First associate a prime number $p_i$ with each
atom type $i = 1,\dots n_{\alpha}$.  Each species $\mathbb{X}$ is made
up of a collection of atom types; let $\rho_{\mathbb{X}}$ be the
product of prime numbers corresponding to each type of atom making up
species $\mathbb{X}$. For example, if elements H and O correspond to
the prime numbers 2 and 3, then $\rho_\h = 2$ and $\rho_{\hho} =
6$. In effect, this yields a prime number representation of each
species where multiplicity is ignored.

Next, the columns of $S$ correspond to chemical reactants and the
rows to chemical products. The entry $s_{ij}$ controls how many
atoms move from species $j$---a sort of reactant---to species $i$---a sort of
product. The operator can only move a positive
amount of species $\mathbb{X}_j$ to $\mathbb{X}_i$ if the former contains all
elements that comprise the latter. If not, then
the $\text{gcd}(\rho_{\mathbb{X}_i}, \rho_{\mathbb{X}_j}) <
\rho_{\mathbb{X}_i}$. But in this case there can be no flow of atoms from
$\mathbb{X}_j$ to $\mathbb{X}_i$, and thus $s_{ij} \equiv 0$.

This technique can also be used to count the total number of entries that are
identically zero in the matrix. Call this total number $\Omega$ and, for
$i=1,\dots,n_S$, let $\lambda_i$ be the number of species $\mathbb{X}_j$ such
that $\gcd(\rho_{\mathbb{X}_i}, \rho_{\mathbb{X}_j}) < \rho_{\mathbb{X}_i}$.  By
the argument in the previous paragraph, $\lambda_i$ is the number of zeros in
the $i$th column of $S$.  Then the number of zeros in $S$ is $\Omega = \sum_{i =
1}^{n_S} \lambda_i$ because the sum is taken with respect to the different species,
and each of these correspond to a different column of $S$.

\subsection{Non-positivity of eigenvalues}
Enforcing the two constraints--- (I) conservation of atoms and (II)
non-negativity of concentrations yields some interesting properties of the
random matrix $S$.  The non-positivity of eigenvalues is consistent with the
constraints: no species can grow arbitrarily large over time.
The proof follows:
\begin{theorem} Let $S$ be any random matrix such that $ES = 0$ and the off-diagonal
elements of $S$ be non-negative. Then (a) the columns sum to zero, (b) the diagonal
is negative, (c) the matrix is weakly diagonally dominant, and (d) the
eigenvalues are non-positive. \end{theorem}

\begin{proof} (a) Consider $ES_{(\cdot,j)} = 0$, where $S_{(\cdot,j)}$ is the $j$th column of $S$. There
are $n_{\alpha}$ equations:
\begin{align*}
  e_{1,1}s_{1,j} + e_{1,2} s_{2,j} + \dots +  e_{1,n_S}s_{n_S,j} &= 0\\
  e_{2,1}s_{1,j} + e_{2,2} s_{2,j} + \dots +  e_{2,n_S}s_{n_S,j} &=0\\
  &\vdots\\
  e_{n_{\alpha},1}s_{1,j} + e_{n_{\alpha}, 2} s_{2,j} + \dots +  e_{n_{\alpha},n_P}s_{n_S,j} &=0.
 \end{align*}
Now add the lines together:
\begin{equation}
\sum_i{ \sum_k{e_{k,i}} s_{i,j}} = 0,
\label{eq:colsum0} \end{equation} but $\sum_k{e_{k,i}} = 1$ by definition. Thus,
\begin{equation}
\sum_i{s_{i,j}} =0.
\label{eq:colsum} \end{equation}\\

(b) In equation (\ref{eq:colsum}), move the diagonal term to the RHS:
\begin{equation}
\sum_{i \neq j}s_{i,j} = -s_{j,j}.
\label{eq:negdiag}
\end{equation} 
Since all off-diagonal terms are non-negative, it must be that the diagonal element
is negative.\\

(c) The line above also shows weak diagonal dominance, since
\begin{align}
|s_{j,j}| &= \left|\sum_{i \neq j}s_{i,j}\right|\\
         &= \sum_{i \neq j} | s_{i,j}|,
\label{eq:diagdom}
\end{align} 
where the second equality holds because all off-diagonal elements are
non-negative.\\

(d) Since $S$ and $S^T$ have the same eigenvalues, we will show that the claim is
true for $S^T$.  Let $B = S^T$ and $B_i = \sum_{j\neq i} | b_{i,j}| =\sum_{j\neq
i} b_{i,j}$ be the sum of off-diagonals in the $i$th row.  Now let $D(b_{i,i},
B_i)$ be the closed disc centered at $b_{i,i}$ with radius $B_i$.  Then the
Gershgorin theorem states that every eigenvalue of $B$ lies within at least one of
the discs \cite{bell65}.  In this case, we have $b_{i,i} = s_{i,i}$ and $B_i =
|s_{i,i}|$, so every eigenvalue lies within at least one disc $D(s_{i,i},
|s_{i,i}|)$, where $s_{i,i} \leq 0$.
\end{proof}

\section{Transform from $\bm{p}$ to $\bm{\xi}$}\label{app:xi}

\begin{table}\begin{centering} \begin{tabular}{r  l}
$\xi_i$\quad= & $p_{j,k}$\\ \hline
$\xi_1$\quad= & $p_{1,1}$\\ 
$\xi_2$\quad= & $-(p_{1,3}+p_{3,3})$\\
$\xi_3$\quad= & $p_{1,5}$\\
$\xi_4$\quad= & $p_{1,7}$\\
$\xi_5$\quad= & $p_{1,8}$\\
$\xi_6$\quad= & $p_{1,9}$\\
$\xi_7$\quad= & $p_{2,2}$\\
$\xi_8$\quad= & $-(p_{2,4}+p_{4,4})$\\
$\xi_9$\quad= & $p_{2,6}$\\
$\xi_{10}$\quad= & $p_{2,7}$\\ 
$\xi_{11}$\quad= & $p_{2,8}$\\ 
$\xi_{12}$\quad= & $p_{2,9}$\\ 
$\xi_{13}$\quad= & $p_{3,1}$\\ 
$\xi_{14}$\quad= & $p_{3,3}$\\ 
$\xi_{15}$\quad= & $-(p_{3,5} +p_{1,5})$\\ 
$\xi_{16}$\quad= & $p_{3,7}$\\ 
$\xi_{17}$\quad= & $p_{3,8}$\\ 
$\xi_{18}$\quad= & $p_{3,9}$\\ 
$\xi_{19}$\quad= & $p_{4,2}$\\ 
$\xi_{20}$\quad= & $p_{4,4}$\\ 
$\xi_{21}$\quad= & $-(p_{4,6} + p_{2,6})$\\ 
$\xi_{22}$\quad= & $p_{4,7}$\\ 
$\xi_{23}$\quad= & $p_{4,8}$\\ 
$\xi_{24}$\quad= & $p_{4,9}$\\ 
$\xi_{25}$\quad= & $-p_{5,7} - 2(p_{1,7} + p_{2,7} + p_{3,7} + p_{4,7} +
(2/3)p_{6,7} + (2/3)p_{7,7})$\\ 
$\xi_{26}$\quad= & $p_{5,8}$\\ 
$\xi_{27}$\quad= & $p_{5,9}$\\ 
$\xi_{28}$\quad= & $p_{6,7}$\\
$\xi_{29}$\quad= & $-p_{6,8} - 3(p_{1,8} + p_{2,8} + p_{3,8} + p_{4,8} +
(1/2)p_{5,8} + (2/3)p_{7,8})$\\ 
$\xi_{30}$\quad= & $p_{6,9}$\\ 
$\xi_{31}$\quad= & $p_{7,7}$\\
$\xi_{32}$\quad= & $p_{7,8}$\\ 
$\xi_{33}$\quad= & $-p_{7,9} - 3(p_{1,9} + p_{2,9} + p_{3,9} + p_{4,9} +
(1/2)p_{5,9} + (2/3)p_{6,9})$\\ 
\end{tabular}
\caption{The transformed variables $\bm{\xi}$ for the H$_2$/O$_2$ operator. \label{tab:xi2}}
\end{centering}
\end{table}

\end{appendices}

\bibliographystyle{siamplain}
\bibliography{rebeccam}
\end{document}